%% file: capm.tex
\documentclass[a4paper,draft=false,DIV=classic,fontsize=10pt]{scrartcl}

\usepackage{dsfont}
\KOMAoptions{DIV=last}

\input{mathheadenglish}

\usepackage{caption}
\begin{document}
\title{Continuous Equilibrium in Affine and Information-Based Capital Asset Pricing Models}
%\subtitle{--                                   A Quantitative Approach for Affine and Information-Based Frameworks --}

\author{{\large Ulrich Horst}\thanks{
Humboldt-Universit\"at zu Berlin, Unter den Linden 6, 10099 Berlin, Germany.}
\e\, \large{ Michael Kupper}{\Large $^\ast$}
\e\, {\large Andrea Macrina}\thanks{University College London, 25 Gower Street, London WC1E 6BT, U.K..}
\e\, \large{Christoph Mainberger}{\Large$^\ast$}
}
\date{October 19, 2012}
\maketitle
\maketitle

\begin{abstract}
\thispagestyle{empty} \noindent
We consider a class of generalized capital asset pricing models in continuous time with a finite number of agents and 
tradable securities. The securities may not be sufficient to span all sources of uncertainty.% 
~If the agents have exponential utility functions and the individual 
endowments are spanned by the securities, 
an equilibrium exists and the agents' optimal 
trading strategies are constant.%
~Affine processes, and the theory of information-based asset pricing 
are used to model the endogenous asset price dynamics and the terminal payoff.%
~The derived semi-explicit pricing formulae are applied to numerically analyze the 
impact of the agents' risk aversion on the implied volatility of simultaneously-traded European-style options.\footnote{The authors are indebted to 
Julio Backhoff for his assistance with the numerical illustrations; special thanks are due to Martin Keller-Ressel 
for helpful comments and fruitful discussions and to an anonymous referee for careful reading and constructive suggestions.%
~U.~Horst and M.~Kupper acknowledge support from the DFG research center MATHEON;~U.~Horst and C.~Mainberger acknowledge
support from the SFB 649 "Economic Risk".%
~A.~Macrina acknowledges generous hospitality by the Humboldt Universit\"at zu Berlin.%
~This paper was written while A.~Macrina was a member of the Department of Mathematics, King's College London.%
~This paper has been presented under the title "Continuous Equilibrium under Base Preferences and Attainable Initial Endowments" 
at the CRC 649 annual meeting (July 2011) and the 
4th Int.~Conference of Math.~in Finance, Kruger National Park (August 2011).% 
~E-mail addresses for correspondence: horst@math.hu-berlin.de; kupper@math.hu-berlin.de; 
a.macrina@ucl.ac.uk; mainberg@math.hu-berlin.de}  
\\


%{\bf AMS Subject Classification}: \red{Primary (...), (...)}

\end{abstract}

%\tableofcontents

\input{introduction}

\input{chap01}
\input{chap022}
\input{chap021}
\begin{appendix}
\input{appendix01}

 \end{appendix}
\bibliographystyle{abbrvnat}
\bibliography{bibliography}
\end{document}

%% file: mathheadenglish.tex
\usepackage{times}
\setkomafont{title}{\normalfont}
\setkomafont{disposition}{\normalfont\normalcolor\bfseries}

\usepackage[english]{babel}
\usepackage[T1]{fontenc}
\usepackage[latin1]{inputenc}
\usepackage[numbers]{natbib}
\usepackage{eurosym}
\usepackage{textcomp}
\usepackage{mathrsfs}
\usepackage{amsfonts}
\usepackage{amssymb}

\usepackage{varioref}

\usepackage{enumitem}
\usepackage[usenames]{color}
\usepackage{graphicx}
\usepackage{subfigure}
\usepackage{stmaryrd}

\newcommand{\ak}[1]{\iffalse #1 \fi}

\usepackage[intlimits]{amsmath}
\usepackage[hyperref,amsmath,thmmarks]{ntheorem}

\usepackage[pdfborder={0 0 0},breaklinks=true]{hyperref}

\theoremseparator{.}
\newtheorem{thm}{Theorem}[section]
\newtheorem{cor}[thm]{Corollary}

\newtheorem{prop}[thm]{Proposition}

\theorembodyfont{\upshape}
\newtheorem{defn}[thm]{Definition}

\theoremsymbol{\ensuremath{\lozenge}}

\theoremsymbol{\ensuremath{\blacklozenge}}
\theoremheaderfont{\itshape}

\theoremsymbol{\ensuremath{\square}}
\theoremheaderfont{\itshape}
\theoremstyle{nonumberplain}

\numberwithin{equation}{section}

\newcommand{\abs}[1]{\left\vert#1\right\vert}

\newcommand{\R}{\mathbb{R}}

\newcommand{\C}{\mathbb{C}}

\newcommand{\E}[1]{{E}\left[#1\right]}

\newcommand{\EQ}[1]{{E}_{Q}\left[#1\right]}
\newcommand{\EFQ}[2]{{E}_{Q}\left[#1 \mid \F_#2 \right]}

\newcommand{\EF}[2]{\E{#1 \mid \F_{#2}}}
\newcommand{\mE}{\mathcal{E}}
\newcommand{\D}{\mathcal{D}}

\newcommand{\U}{\mathcal{U}}

\newcommand{\F}{\mathcal{F}}

\renewcommand{\mid}{\;|\;}

\newcommand{\ef}[1]{{\rm e}^{#1}}

\DeclareMathOperator*{\argmax}{arg\,max}

\newcommand{\ba}{\begin{align}}
\newcommand{\ea}{\end{align}}

\newcommand{\bas}{\begin{align*}}
\newcommand{\Als}{\end{align*}}

\newcommand{\be}{\begin{equation}}
\newcommand{\ee}{\end{equation}}

\newcommand{\bes}{\begin{equation*}}
\newcommand{\ees}{\end{equation*}}

\newcommand{\bea}{\begin{eqnarray}}
\newcommand{\eea}{\end{eqnarray}}

\newcommand{\beas}{\begin{eqnarray*}}
\newcommand{\eeas}{\end{eqnarray*}}

\newcommand{\q}{\quad}
\newcommand{\e}{\enspace}
\newcommand{\qq}{\qquad}

\newcommand{\brak}[1]{\left(#1\right)}    % round brackets
\newcommand{\crl}[1]{\left\{#1\right\}}   % curly brackets
\newcommand{\edg}[1]{\left[#1\right]}     % edgy brackets

\newcommand{\hv}{\hat\vartheta}

\newcommand{\vt}{\vartheta}
\newcommand{\tg}{\tilde\gamma}

\renewcommand{\mid}{\,|\,}

%% file: introduction.tex
\section*{Introduction}
\addcontentsline{toc}{section}{Introduction}
\markboth{\uppercase{Introduction}}{\uppercase{Introduction}}

In this paper we propose an analytically-tractable equilibrium model in continuous time, within which 
financial securities are priced in a generalized capital asset pricing model (CAPM).
\\[5pt]
It is well known that when markets are incomplete, competitive equilibria may fail to exist. 
Even if they exist, they may not be Pareto optimal, nor supportable as equilibria of a suitable representative agent economy. 
The equilibrium analysis of incomplete markets is therefore always confined to special cases, for instance to 
single agent models \cite{HeLeland01, GarleanuEtAl01}, multiple agent models where markets are complete 
in equilibrium \cite{DuffieHuang01, HorstEtAl01, KaratzasEtAl01}, or models with particular classes of goods \cite{JofreEtAl01} 
or preferences \cite{CarmonaEtAl01}.
\\[5pt]
In \cite{chkp}, the authors recently established existence and uniqueness of equilibrium results for incomplete financial 
market models in discrete time when agents' preferences are translation invariant\footnote{Exponential utility functions, for instance,
are translation invariant after a logarithmic transformation.}.%
~In the situation where uncertainty is spanned by finitely many random walks, they showed that the equilibrium dynamics can 
be described as the solution 
to a coupled system of forward-backward stochastic difference equations. The system is usually high-dimensional because one obtains 
one equation per security and market participant. 
This renders simulations and calibrations of the model cumbersome, if not impossible. 
Within the framework of generalized CAPMs, that is, if all agents share the same base preferences 
(as in the case of exponential utility functions) and the endowments lie in the span of the tradable assets, the 
system simplifies to a single equation representing the equilibrium utility of some representative agent. 
Furthermore, the equilibrium price process depends only on the aggregated endowment,
the market risk aversion, and the flow of market information.
It is in this sense that these three items fully characterize equilibrium prices in generalized CAPMs.
\\[5pt]
In this paper we extend the generalized CAPM analyzed in  \cite{chkp} to continuous time
when agents' preferences are of the expected exponential type.% 
~In particular, the advantage of the herewith presented continuous-time framework 
is that we obtain (semi-)explicit formulae for equilibrium prices. 
If not explicitly computable, key equilibrium quantities can be computed using numerical integration only---no Monte Carlo methods are needed.% 
~We consider a model with a finite number of agents, 
which are initially endowed with an attainable random payoff. 
They trade a finite number of securities so as to maximize expected exponential utility from terminal wealth. 
The financial securities are characterized by their terminal payoffs, which we assume to be functions of finitely many market factors.% 
~The market factors may or may not be observable to the agents.%
~Affine processes, and the theory of information-based asset pricing 
are used to model the endogenous asset price dynamics and the terminal payoff.
\\[5pt] 
Within our first approach, the dynamics of the market factors follows an affine process that generates the 
market filtration.% 
~Affine processes are extensively used in mathematical finance (see for instance \cite{DuffieSingleton01, duffieAffine, MKRphd} and references therein), 
as they lend themselves to a transparent mathematical analysis and to the application of efficient numerical methods. 
We show that within an affine framework, equilibrium securities prices are given by the quotient of two integrals. 
Both integrals are the product of an exponential function evaluated at the current state of the factor process 
and the Fourier transform of a smooth function. 
Representing equilibrium prices in terms of deterministic integrals allows for a fast and efficient numerical analysis of other equilibrium quantities, 
such as option implied volatilities. 
We analyze implied volatilities for two single-security benchmark models: 
(i) an additive Heston stochastic volatility model, and (ii) a pure jump Ornstein-Uhlenbeck model. 
Both models reproduce the well-documented smile-effect of implied volatilities and identify investor risk aversion as a 
key determinant of implied volatilities.
\\[5pt]
The second approach to continuous equilibrium presented in this paper is based on the theory of infor-mation-based asset 
pricing, see \cite{andrea04} and \cite{andrea01}.%
~Within this approach, the asset price dynamics is explicitly generated by 
taking the conditional expectation of the future cash flows, which are multiplied 
by the pricing rule, given the partial information about the market factors that is available to the agents. 
The filtration is modeled by stochastic processes, which (i) carry information about 
the a priori distribution of the market factors,
and (ii) embody pure noise preventing market participants from accessing full knowledge 
as to what is the ``true'' value of the asset at any time before the cash flows occur.%
~We use the information-based framework to show the dependence of the equilibrium prices of credit-risky securities on 
information about the financial standing of a company.
\\[5pt]
The paper is structured as follows.%
~A general existence result along with a discussion on the information-generating processes is given in Section \ref{sec01}.%
~In Section \ref{sec022} and \ref{sec021}, we present affine and information-based equilibrium pricing models, respectively.%
~Proofs to the theorems are collected in the appendix.

%% file: chap01.tex
\section{A Generalized Capital Asset Pricing Model}\label{sec01}
We consider an equilibrium model in continuous time with a finite set  $\mathbb A$ of economic agents. 
Uncertainty is modeled by a probability space $(\Omega, \F, P)$ carrying a filtration $(\F_t)_{t\in[0,T]}$. 
The filtration  captures the flow of information that is available to the agents over the trading period $[0,T]$,
and is assumed to satisfy the usual assumptions of completion and right-continuity. 
In what follows, all equalities and inequalities are to be understood in the $P$-almost sure sense.

\subsection{Existence of Equilibrium}
The agents can lend to and borrow from the money market account at some exogenously given interest rate, and they can trade $K$ securities. 
The securities are in net supply $n = (n^1,\dots,n^K)\in {\R^K}$ and characterized by their terminal payoffs $S_T=(S^1_T,\ldots,S^K_T)$, 
which we assume to be $\F_T$-measurable random variables. Securities are priced to match demand and supply.% 
~Each agent $a \in {\mathbb A}$ is initially endowed with some $\F_T$-measurable random payoff $H^a$ of the form 
\begin{equation*}%\label{endowa} 
H^a = c^a + \eta^a\cdot S_T\,,
\end{equation*}%nd{align} 
for constants $c^a \in \R$ and $\eta^a \in \R^K$.~Furthermore, at each time $t\in[0,T]$ the agent's preferences can be described by the utility functional 
\begin{equation*}%\label{eq101}
U^a_t(X) = -\frac{1}{\gamma^a}\log \brak{\EF{ \ef{-\gamma^a X}}{t}}\,,%\e,\e X\e \mathcal F_T\mbox{-measurable}\,,
\end{equation*}
where $\gamma^a>0$ is the risk aversion parameter. Thus at time $t \in [0,T]$, the agent faces the optimization problem  
\begin{equation*}%\label{eq11}
\sup_{\vt \in \Theta} U^a_t\brak{H^a + \int_t^T\vt_u dS_u }\,,
\end{equation*}
where the set of admissible trading strategies $\Theta$  %for agent $a$, 
is given by
\begin{equation*}%\label{admissibility}
\Theta=\Big\{\vt\in L(S) \,:\, 
G(\vt)\,\text{is a $\tilde Q$-supermartingale, for all $\tilde Q\in\mathcal P$}\Big\}\,.
\end{equation*}
Here, $L(S)$ and $G_t(\vt):=\int_0^t \vt_u dS_u$ denote the set of $S$-integrable predictable processes and the gains process, respectively,
whereas $\mathcal P$ denotes the set of all equivalent martingale measures (EMM) for $S$.\footnote{Note that in equilibrium, there is an 
EMM $Q$, that is, an equivalent probability measure $Q$ under which the price process $S$ will be a true martingale.%
~In particular, $S$ will be a $P$-semimartingale.%
~For related discussions on suitable sets of admissible strategies 
see for instance \cite{Delbaen2006}, \cite{SchweitzerEtAl}, or \cite{biaginifritelli1}.}
\\[5pt]
The goal is now to establish existence of a (discounted) equilibrium price 
process $(S_t)_{t \in [0,T]}$.\footnote{For simplicity, we assume that the trading horizon $T$ is short so 
that interest rate risk can be ignored.} Since all agents share the same base preferences, and because 
all payoffs lie in the span of the tradable assets, our model can be viewed as a generalized CAPM. 
Just like in the classical CAPM, in our incomplete market model existence of an equilibrium can be established using 
the standard representative agent approach that underlies equilibrium models of complete markets.%
~Furthermore, all agents share the market portfolio according to their risk aversion in equilibrium.% 
~The equilibrium pricing kernel depends on the agents' preferences and endowments, however only through 
the endowment- and supply-adjusted risk aversion
\begin{equation}\label{revi02}
	\tilde{\gamma} := \gamma(\eta + n) \in \R^K \,.
\end{equation}
Here, $\eta := \sum_{a} \eta^a$ denotes the aggregate endowment and $\gamma^{-1}:=\sum_{a} \gamma_a^{-1}$ can be viewed as the market risk aversion.  
The following result can be proved by standard duality results for entropic utility 
functions; see \citep[Theorem 5.1]{chkp} for a related result in discrete time. 
\begin{thm}\label{density}
Suppose that the following integrability conditions hold:
\begin{align}\label{equiS1}
	\exp \brak{ -\tg\cdot S_T } \in L^1(P) \q \mbox{and} \q S_T\in L^1(Q)^K\,,
\end{align}
where $Q$ is an equivalent probability measure with density
\begin{align}\label{density2}
\frac{dQ}{dP} = \frac{\exp(-\tilde{\gamma} \cdot S_T)}{\E{\exp(-\tilde{\gamma}\cdot S_T)}}\,.
\end{align} 
Then, the price process $S$ defined by
\begin{equation}\label{equiS}
S_t=\EFQ{S_T}{t}\,,\q t\in[0,T]\,,
\end{equation}%nd{align} 
together with  the constant trading strategies %in the securities%$\hat\vartheta_t^a$ %is constant and equals
\begin{equation*}%\label{equiS4} 
\hat\vt_t^a \equiv \frac{\gamma}{\gamma^a}(n+\eta)-\eta^a, \q a \in \mathbb{A},
\end{equation*} 
constitutes an equilibrium.
\end{thm} 
We notice that the equilibrium pricing kernel $Q$ depends only on the terminal payoffs weighted by the endowment- and supply-adjusted risk aversion. 
In particular, if the $k$-th security is in zero endowment-adjusted supply, 
that is, if $\eta^k + n^k=0$, then its payoff does not affect the equilibrium pricing kernel.
\\[5pt]
Furthermore, the integrability assumption on $S$ under the pricing measure $Q$ guarantees that equilibrium prices are $Q$-martingales. 
Hence they are, by \eqref{equiS1} and \eqref{density2}, $P$-semimartingales and thus well defined as an integrator 
in the sense  of~\citep[Chapter II and IV]{protter01}.

\subsection{The Market Filtration}

The previous theorem established existence of a continuous equilibrium under no assumptions on the underlying filtration $(\F_t)$.%
~We emphasize that the construction of the filtration characterizes the dynamics of the derived price processes.%
~In order to obtain (semi-)explicit equilibrium price processes, we assume
 that the terminal payoffs depend in a functional form on a vector $X$ of market factors the distribution of which is known to the agents.%
~We define the following:
\[
	S^k_T = f^k(X)\,.
\] 
We assume that the market filtration $(\F_t)_{t\in[0,T]}$, 
to which the equilibrium prices will be adapted, is generated by an observable stochastic process $(\xi_t)$ such that, possibly up to a
constant, $\xi_T = X$.%
~Equilibrium dynamics are then studied within an affine and an information-based framework. 
The first approach assumes that the dynamics of the market factors follow an affine process; 
in the second approach the observables generating the market filtration are modeled by Brownian random bridges
with drift from zero to $X$. 

%% file: chap022.tex
\section{Affine Equilibrium Framework}\label{sec022}
In this section, we assume that the dynamics of the market factors $\xi$ are observable and that they follow an affine process $Y$,
that is $\xi=Y$.%
~After specifying the setup and following a brief introduction into the theory of affine processes, the results in Section \ref{sec01}
are used to derive equilibrium pricing formulae in Section \ref{sec0221}. This is followed by an analysis of  equilibrium option prices  
in Section \ref{sec030}  
and equilibrium asset prices in a Heston stochastic volatility framework and an Ornstein-Uhlenbeck jump model in Sections 
\ref{sec031} and \ref{sec032}, respectively.% 
~Since we consider a linear payoff structure of the underlying asset $S_T=X_T$ from Section \ref{sec030} onwards, 
negative equilibrium prices can not be excluded a priori.%
~However, this can be avoided by %either considering only short trading horizons $T$, because
 either directly modeling the log-payoff of the underlying, that is $S_T=\exp(X_T)$, or, as in the present work,
considering only short trading horizons $T$.%
~In this case, option prices obtained from a model 
and its ``logarithmic counterpart" are quite close, compare for instance the discussion in \cite{SmayerTmann01}.%
~We choose the first approach, since the verification of the integrability conditions in Theorem \ref{MAINTHM} is 
more involved in the case of a log-payoff.%
~The additional challenge is due to the ``double exponential" structure.%
%We adhere to the first approach, mainly because for the case of modeling the log-payoff a verification of the integrability 
%assumptions in our main Theorem \ref{MAINTHM} turns out to be more cumbersome due to a "double-exponential" structure.
~We emphasize however that, once this is achieved, all our results can be adapted and hence extended also to longer trading horizons.

\subsection{Setup and Equilibrium Pricing Formulae}\label{sec0221}

In this section, we consider the case where the payoff $S_T$ is a functional of an observable affine factor process. 
To this end, we assume that the underlying probability space $(\Omega,\F, P)$ is rich enough to support 
an affine Markov process $Y$ taking values in the state space $D:=\R_+^{m}\times\R^{n}$. 
\\[5pt]
We set $d=m+n$ and write $Y=(V,X)$. We interpret $X \in \R^n$ as the factor process that determines the payoff and $V \in \R^m_+$ 
as a process driving it; a typical example would be a stochastic volatility model. We assume that $Y^T$, the Markov process 
stopped at time $T$, is conservative, meaning that  there are no explosions or absorbing states up to time $T$. 
The market filtration $(\F_t)_{t\in[0,T]}$ is then chosen to be the one generated by $Y$: 
\begin{equation*}%\label{mfilt}
\F_t = \sigma(Y_s\,, s\le t), \quad t\in[0,T].
\end{equation*}
Usually, one associates with $Y$ a family of probability measures $(P^y)_{y \in D}$, which represents the law of 
the process $Y$ starting at $y\in D$.%\footnote{That is, $Y_0=x$, $P^x$-almost surely.}, 
~Since every affine process is a Feller process, the filtration $(\F_t)$ can be completed with respect to the family $(P^y)_{y \in D}$ so that
the filtration is
automatically right-continuous \citep[Section III.2]{Revuz1999}.

\subsubsection{Affine processes}\label{ap}
Before turning to the problem of equilibrium pricing, we recall some useful results on 
affine processes, the details of which can be found in \cite{duffieAffine} or in \cite{MKRphd}. 

\begin{defn}\label{defaffreg}
An affine process is a stochastically continuous\footnote{A stochastic process $Y$ is stochastically continuous, 
if for any sequence $(t_m) \rightarrow t$ in
$\R_+$, $Y_{t_m}$ converges to $Y_t$ in probability.},
time-homogeneous Markov process ($Y, P^y)$ %on our filtered probability space $(\Omega,\mf, \mf_T)$ 
with state-space $D$, of which log-characteristic function is an affine function of the state vector.%
~That is, there exist functions $\phi:\R_+\times i \R^{d} 
\to \C$ and $\psi:\R_+\times i \R^{d} \to \C^{d}$ such that
\begin{equation}\label{affinedef1}
E^{y}\edg{\exp\brak{u\cdot Y_t}}= \exp\edg{\phi(t,u) +\psi(t,u)\cdot y}\,, 
\end{equation}%nd{align}
for all $y \in D$ and $(t,u) \in \R_+\times i \R^{d}$.
An affine process $Y$ is called regular,
if the derivatives
\begin{equation*}
F(u) :=  \left.\partial_t \phi(t,u)\right|_{t=0^+} \q,\q \left. R(u) := \partial_t \psi(t,u)\right|_{t=0^+}
\end{equation*}%nd{align*}
exist for all $u \in \U := \crl{u = (u_v,u_x) \in \C^m\times\C^n \,:\,{\rm Re}(u_v) \le 0, \,\,{\rm Re}(u_x)=0 }$ and are continuous 
in $u=0$.\footnote{In 
the recent work \cite{MKRaffreg}, the authors actually show that each
affine process as defined above is regular, whereas in \cite{duffieAffine} and \cite{MKRphd} regularity is still an assumption on $Y$.}
\end{defn}
The definition of an affine process $Y$ implies that the $\F_t$-conditional characteristic function of $Y_T$ $(T \geq t)$ is an 
affine function of $Y_t$: 
\begin{align}\label{affinedef2}
\EF{\exp\brak{u\cdot Y_T}}{t} = \exp\edg{\phi(\tau,u) + \psi(\tau,u)\cdot Y_t}\,,
\end{align}
for all $(\tau,u) \in \R_+\times i \R^{d}$, where $\tau := T-t$. The affine property will be used in this form throughout. 

The admissible parameters associated with an affine process $Y$ determine its generator and its functional characteristics $F$ and $R$.% 
~The functional characteristics completely determine a regular affine process, since the functions $\phi$ and $\psi$ 
satisfy generalized Riccati equations of the form $\partial_t \phi(t,u) = F(\psi(t,u))$ and $\partial_t\psi(t,u)=R(\psi(t,u))$; see 
Appendix \ref{appendix011} for further details. 
\\[5pt]
Although the special form of the log-characteristic function of an affine process perfectly lends itself to tractable 
computations, we need to consider 
a class of processes for which formulae \eqref{affinedef1} or \eqref{affinedef2} extend to a broader subspace of $\C^d$ than $i\R^d$.\footnote{By 
extension it is meant that the functions $\phi$ and $\psi$ can be uniquely analytically extended to a suitable subspace 
of $\R_+\times\C^d$.}
\ak{
\begin{definition}\label{defanalytic}
For $Y$ a regular affine process, and for each $t\ge 0$, the following sets are considered:
\begin{align}\label{defanalytic1}
\D_t \,\,&:=\crl{ z \in \R^d \,: \, \sup_{0\le s\le t} E^y\edg{\exp(z\cdot Y_s)} < \infty\,, \e\text{for all}\e y \in D  },\nonumber\\
\D_{t^+} &:= \bigcup_{s>t}\D_s \q\mbox{and}\q \D := {\rm int}\,\D_{0^+}\cup \{0\}.
\end{align}
$\D$ is called the real domain of $Y$, and $Y$ is said to be analytic, if the interior $\textrm{int}\,\D$ of $\D$ is non-empty.
\end{definition}
}
%\begin{remark}
It is shown in \citep[Chapter 3]{MKRphd} that the functions $\phi$ and $\psi$ 
characterizing the process $Y$ have unique extensions to analytic functions on the interior $\textrm{int}\,\mE_{\C}$ of the tube domain 
%$\mE_{\C}$ defined by
%\begin{align*}
$\mE_{\C} := \crl{ (t,u) \in \R_+\times \C^d : (t,{\rm Re}(u)) \in \mE }$,
%\mE_{\C} :=& \crl{ (t,u) \in \R_+\times \C^d\,:\, (t,{\rm Re}(u)) \in \mE }\\
%\text{where}\e\, \mE :=& \,\{(t,v) \in \R_+\times  \R^{d} \,:\, v \in \D_{t^+}\}\,.
where $\mE := \{(t,v) \in \R_+\times  \R^{d} : v \in \D_{t+}\}$ and the set $\D_{t+}$ is defined by
$\D_{t+} := \bigcup_{s>t}\{ z \in \R^d :  \sup_{0\le r\le s} E^y\edg{\exp(z\cdot Y_r)} < \infty\,, \e\text{for all}\e y \in D  \}$.
%\end{align*}
The extensions still satisfy the aforementioned Riccati equations and %the affine transformation formulas 
\eqref{affinedef1} and \eqref{affinedef2} 
extend to $\mE_{\C}$.\footnote{More precisely, \citep[Lemma 3.12]{MKRphd} states that this holds on the 
set $\{(t,u) \in \mE_{\C}\,:\, \abs{E^0\edg{\exp(u\cdot Y_s)}} \neq 0\,, \e\text{for all $s \in [0,t)$}\}$, whereas \citep[Lemma 3.19]{MKRphd} then 
yields that both sets coincide.}%
%~The regular affine process $Y$ is said to be analytic, if the interior $\textrm{int}\,\D_{0+}$ is non-empty.%
~Recently, an alternative characterization of the extensibility of the affine transform formula \eqref{affinedef2} has been given in \cite{MKRMayer}.
%\end{remark}

\subsubsection{Equilibrium pricing formulae}

We are now ready to state the main result of this section, that is a semi-explicit formula for 
the equilibrium price processes in an affine framework. 
For simplicity, we restrict the analysis to processes $Y=(V,X)$ with state space $D=\R_+ \times \R$, and we assume that 
the agents can trade $K$ securities $S^1,\dots,S^K$ with terminal payoffs 
\begin{equation} \label{payoff}
	S^k_T = f^k(X_T)\,,
\end{equation}
for payoff functions $f^k: \R \to \R$. Under suitable integrability conditions our results carry over to more general 
payoff functions of the form $f^k(Y_T)$ and to affine processes on multi-dimensional state spaces. However, the resulting pricing formulae 
would be quite cumbersome and the Riccati equations that determine the processes' functional characteristics would no longer be 
solvable in closed form (the semi-explicit structure of the solution would be preserved, though).%
~We define $f(x) := (f^1(x),\dots,f^K(x))$. 

\begin{thm}\label{MAINTHM}
Let $Y = (V,X)$ be an affine process on $\R_+\times\R$, and suppose that the terminal payoffs of the securities are of the form (\ref{payoff}). 
Suppose furthermore that there exists a vector of damping parameters $(\alpha^1,\ldots,\alpha^K,\beta)\in\R^{K+1}$ such that
the functions 
\begin{align} 
g_\zeta^k(x) &:= \exp\brak{\alpha^k x} f^k(x)\exp\brak{-\zeta\cdot f(x)}\,,\label{MAINTHM2}\\
h_\zeta(x) &:= \exp\brak{\beta x} \exp\brak{-\zeta\cdot f(x)}\,\label{MAINTHM3} \,,
\end{align}
and their respective Fourier transforms, 
\begin{equation*}%\label{newidea2}
 \hat g^k_\zeta(s) = \int_{\R} {\rm e}^{-isy} g^k_\zeta(y) dy\qq\mbox{and}\qq
 \hat h_\zeta(s) = \int_{\R} \ef{-isy} h_\zeta(y) dy\,,
\end{equation*}%nd{align}
are integrable for all $\zeta$ in some neighbourhood of $\tg$, and that
\begin{equation}\label{MAINTHM4} \brak{T,(0,-\alpha^k)} \in \mE\,,\q\text{for all $k$,} \q\text{and} \q \brak{T,(0,-\beta)} \in \mE. \end{equation}
Then, with $\hat g^k(s) \equiv \hat g^k_{\tg}(s)$ and $\hat h(s) \equiv \hat h_{\tg}(s)$, the following holds:
\begin{itemize}
\item[(i)] The equilibrium price of $S$ at time $t$ is a function of $\tau:=T-t$ and the current state of the process $Y$, and 
the price of the $k$-th security at time $t\in[0,T]$ is given by
\begin{equation}\label{mainTHMform}
S^k_t = \frac{\int_{\R} \exp\edg{\phi\big(\tau, (0,-\alpha^k+is)\big) + \psi\big(\tau, (0,-\alpha^k+is)\big) \cdot Y_t}\,\hat g^k(s) \,ds}
{\int_{\R} \exp\edg{\phi\big(\tau, (0,-\beta+is)\big) + \psi\big(\tau, (0,-\beta+is)\big) \cdot Y_t}\,\hat h(s) \,ds}\,.
\end{equation}
Here, $\phi$ and $\psi$ denote the analytic extensions of the functions introduced in Definition \ref{defaffreg}.

\item[(ii)] The equilibrium price process of $S$ at time $t$ can alternatively be computed by 
\begin{equation}\label{AffPropEQ}
S^k_t = \left. - \frac{\partial}{\partial \zeta^k} \,H(\zeta) \big/ H(\tg) \,\right|_{\zeta=\tg}.
\end{equation}
Here, the function $H: \R^K \to \R$ is given by
\begin{equation*}
H(\zeta)=\frac{1}{2\pi}\int_{\R} \exp\left[\phi\Big( \tau,\brak{0,-\beta +is} \Big) + \right.
\left.\psi\Big( \tau, \brak{0,-\beta + is} \Big) \cdot Y_t\right]\,\hat h_\zeta(s)\,\,ds \,.
\end{equation*}  
%where we %allow for a functional dependence of $\beta\equiv\beta(\tg)$ on $\tg$, and 
%used the explicit dependence of the function $h(s) \equiv h_{\tg}(s)$ on 
%$\tg$ through \eqref{MAINTHM3}.
\end{itemize}
\end{thm}
The benchmark case where only one security is in non-zero endowment-ad-justed supply and its payoff function is linear, and all other securities 
are in zero endowment-adjusted supply, does not require Fourier transform methods, %in order to price the non-zero endowment-adjusted supply security, 
as shown by the following corollary.

\begin{cor}\label{AffPropCor}
Let the process $Y$ and the functions $\hat g^k(s)$ be as in Theorem \ref{MAINTHM}.~Let us further assume that there is only
one security, denoted by $S^1$, in non-zero endowment-adjusted supply, that is 
\begin{equation*} \tg = \brak{\gamma(\eta^1+n^1),0,\ldots,0}\,.\end{equation*} 
If furthermore $S^1_T = X_T$ and $\tg$ satisfies $\brak{T,(0,-\tg^1)} \in \mE$,
then the equilibrium price process of $S^1$ is given by 
\begin{equation}\label{AffPropCorEQ}
S^1_t =  
\left.\big[ \partial_{u_x} \phi(\tau,u) + \partial_{u_x} \psi(\tau,u)\cdot Y_t\big]\,\,\right|_{u=(0,-\tg^1)}\,,\q t\in[0,T]\,,
\end{equation}
where $\tau:=T-t$ and $\partial_{u_x}$ denotes the partial derivative with respect to the second argument of the vector $u=(u_v,u_x)$.% 
~Furthermore, whenever the remaining securities $(S^2,\dots,S^K)$ satisfy the assumptions of Theorem \ref{MAINTHM}, their price processes
equal
\begin{equation}\label{AffPropCorEQa}
S^k_t=\frac{1}{2\pi} \int_\R \exp\big[  \Delta^{\alpha^k,\tg^1}_\tau(\phi) 
\e +\e \Delta^{\alpha^k,\tg^1}_\tau(\psi)\cdot Y_t  \big]\,\hat g^k(s)\, ds\,,
\end{equation}
for $k=2,\dots,K$, and each $t\in[0,T]$.% 
~The shift operator $\Delta^{w,z}_t(\varphi)$ in \eqref{AffPropCorEQa} is defined by 
\[
	\Delta^{w,z}_t (\varphi) :=  \varphi\big(t,(0,-w+is)\big) - \varphi\big(t,(0,-z)\big).
\]
\end{cor}
\input{chap30}

\input{chap31}

\input{chap32}

%% file: chap30.tex
\subsection{Pricing of Call Options}\label{sec030}
We are now going to establish semi-explicit pricing formulae for European call options.%
~The main challenge will be to find suitable ``damping'' functions such that the Fourier methods of Theorem \ref{MAINTHM} can be 
applied. 
Specifically, we consider a market model with a single stock with terminal payoff $S_T=X_T$ and $N$ call 
options on the stock with payoffs $C^i_T=(S_T - K_i)^+$, for $i=1,\ldots, N$, and strike prices $K_1 <\ldots< K_N$. 
The stock and the options are traded simultaneously and hence collectively influence the equilibrium pricing kernel.
The flattening functions for $S$ and $C^k$ are denoted $\alpha$ and $\alpha^k$, respectively;
the corresponding weighted payoff functions are denoted $g$ and $g^k$, respectively.
We first state the  pricing formula for the most general case of multiple simultaneously traded options in non-zero endowment-adjusted supply. 
The formulae are a direct application of Theorem \ref{MAINTHM}. 
Subsequently, we consider the %particularly transparent 
cases where either a single option in non-zero endowment-adjusted supply is traded,
or multiple options in zero endowment-adjusted supply are traded.

\subsubsection{Multiple, simultaneously traded options}\label{sec221} 
Let us first consider the general case where $N>0$ call options and one stock in %possibly 
non-zero endowment-adjusted supply are traded.%
~As an illustration, 
we assume that throughout Sections \ref{sec221} and \ref{sec222}
all supply-adjusted risk aversion parameters satisfy
\begin{equation*}%\label{sarv2}
\tg^1=\ldots=\tg^{N+1}=\gamma\,.
\end{equation*}
The pricing measure is then given by 
\begin{align}\label{multidensity}
\frac{dQ}{dP} = \frac{\exp\brak{-\gamma\brak{S_T + \sum_{i=1}^N(S_T-K_i)^+}}}{\E{\exp\brak{-\gamma\brak{S_T+\sum_{i=1}^N(S_T-K_i)^+}}}}\,,
\end{align}
and the following result is an immediate consequence of Theorem \ref{MAINTHM}. 
\begin{thm}\label{ThmEx1}
Given that $\alpha$ and $\beta$ satisfy $\gamma<\alpha,\beta<(N+1)\gamma$ and $\eqref{MAINTHM4}$, the equilibrium price of the underlying security $S$ at 
time $t\in[0,T]$ is given by
\[%\label{eq301}
S_t = \frac{\int_\R \exp\edg{\phi\big(\tau, (0,-\alpha+is)\big) + \psi\big(\tau, (0,-\alpha+is)\big)\cdot Y_t }\,\hat g(s) ds}
{\int_\R \exp\edg{\phi\big(\tau, (0,-\beta+is)\big) +
 \psi\big(\tau, (0,-\beta+is)\big)\cdot Y_t}\,\hat h(s) ds}\,,
\]%ees%nd{align}
and the price of the $k$-th call option is given by
\[%bes%\label{eq302} 
C^k_t = \frac{\int_\R \exp\edg{\phi\big(\tau, (0,is)\big)+ 
\psi\big(\tau, (0,is)\big) \cdot Y_t}\,\hat g^k(s) ds}{\int_\R \exp\edg{\phi\big(\tau, (0,-\beta+is)\big)
 + \psi\big(\tau, (0,-\beta+is)\big) \cdot Y_t}\,\hat h(s) ds}\,,
\]%ees%nd{align}
for $k = 1,\ldots,N$.~Here the functions $\hat g$, $\hat g^k$ and $\hat h$ are given by 
\begin{multline*}%\label{hardFourier}
\hat g(s) = \sum_{j=1}^{N} \exp\brak{\gamma\sum_{k=1}^{j-1}K_k} \, \exp\edg{(-is + \alpha - j\gamma)K_j}
\left[\brak{ \frac{-K_j\gamma}{(-is + \alpha -j\gamma)(-is + \alpha -(j+1)\gamma)} }\right. \\
+\left.\brak{ \frac{1}{(-is + \alpha - (j+1)\gamma)^2} - \frac{1}{(-is + \alpha -j\gamma)^2} } \right]\,, 
\end{multline*}
\begin{equation*}
\hat h(s) =  \sum_{j=1}^{N} \exp\brak{\gamma\sum_{k=1}^{j-1}K_k} \, 
\exp\edg{(-is + \beta -j\gamma) K_j}
 \edg{ \frac{-\gamma}{(-is + \beta - j\gamma)(-is + \beta -(j+1)\gamma)} } \,,
\end{equation*}
\begin{multline*}
\hat g^k(s) = \exp\brak{\gamma\sum_{h=1}^{k-1}K_h} \, \exp\edg{(-is - k\gamma) K_k}\edg{ \frac{1}{(-is - (k+1)\gamma)^2} }  \\
+  \sum_{j=k+1}^{N} \exp\brak{\gamma\sum_{h=1}^{j-1}K_h} \exp\edg{(-is - j\gamma) K_j}
\left[  \brak{ \frac{-(K_j-K_k)\gamma}{(-is - j\gamma)(-is - (j+1)\gamma)} }\right. \\
\left. + \brak{ \frac{1}{(-is - (j+1)\gamma)^2} - \frac{1}{(-is - j\gamma)^2} } \right]\,.
\end{multline*}
\end{thm}
The assumption $\gamma<\alpha,\beta<(N+1)\gamma$ imposed on the damping factors %and such that the functions 
ensures that the functions $g,g^k,h$ of \eqref{MAINTHM2} and \eqref{MAINTHM3} allow for an integrable Fourier transform.%  
~In what follows, all model parameters have to be chosen such that \eqref{MAINTHM4} is satisfied and hence \eqref{affinedef2} applies.%
~Further details are discussed below.

\subsubsection{A single option model}\label{sec222}  

The pricing kernel \eqref{multidensity} and the Fourier transforms from Theorem \ref{ThmEx1} simplify considerably when 
only one option with strike $K>0$ is traded. In this case the price processes $(S_t)$ and $(C_t)$ can be 
computed as in Theorem \ref{ThmEx1} by
\begin{align*}
\hat g(s)&= \exp\edg{(\alpha - \gamma - is)K}\left[\frac{-K\gamma}{(-is - \gamma +\alpha)(-is - 2\gamma + \alpha)}\right.\\ 
&\hspace{4.0cm}\left.+\left(\frac{1}{(\alpha - 2\gamma - is)^2} - \frac{1}{(\alpha - \gamma -is)^2}\right)\right]\,.\\
\hat h(s)&= \exp\edg{(\beta - \gamma - is)K}\brak{\frac{-\gamma}{(\beta - \gamma - is)(\beta - 2\gamma -is)}}\,.\\
\hat g^1(s)&= \exp\edg{-(is+\gamma)K}\frac{1}{(-is - 2\gamma)^2}\, \,.
\end{align*}

\subsubsection{Options in zero-endowment-adjusted supply} 
Let us finally consider the simplest situation in which all options are in zero endowment-adjusted supply. 
In this case, the equilibrium pricing kernel is independent of option payoffs and one only needs to find  
a suitable $\alpha$ corresponding to the weighted payoff function \eqref{MAINTHM2} 
in Theorem \ref{MAINTHM}. 
The simple choice $\alpha=0$ already guarantees that the Fourier-transform 
\begin{equation*}
\hat g^1 (s)= \exp\edg{-(is + \tg^1)K}\frac{1}{(is + \tg^1)^2}
\end{equation*}
of the function $g^1(x) := \ef{-\tg^1 x}(x - K)^+$ is integrable.%
~The price process $S$ is then given by \eqref{AffPropCorEQ}, and the 
price of the call option at time $t\in[0,T]$ is given by
\begin{equation*}%\label{eq303}
C_t = \frac{1}{2\pi} \int_\R \exp \big[ \Delta^{0,\tg^1}_\tau (\phi)  + \Delta^{0,\tg^1}_\tau (\psi_1) V_t  + 
\Delta^{0,\tg^1}_\tau (\psi_2) X_t  \big]\,\hat g^1(s) ds \,,
\end{equation*}
with $\tau:=T-t$ and $\Delta^{0,\tg^1}_\tau$ defined in Corollary \ref{AffPropCor}.

%% file: chap31.tex
\subsection{Equilibrium Dynamics in a Stochastic Volatility Model}\label{sec031}
By choosing the dynamics of $Y$ according to the Heston stochastic volatility model
\cite{Heston}, it is possible to 
derive explicit equilibrium stock price formulae.
Let $Y=(V,X)$ be determined by
\begin{align}
%\begin{split}
dV_t =& \e (\kappa -\lambda V_t)dt + \sigma \sqrt{V_t}dW^1_t &  V_0& = v_0\,, \nonumber\\
dX_t =& \e \mu dt + \e\sqrt{V_t} dW^2_t &  X_0& = x_0\,,\label{hestondyn}
\end{align}
where $(\Omega,\F,P)$ is assumed to be rich enough to support the two-dimensional Brownian motion $W=(W^1,W^2)$.\footnote{The more 
general case of correlated Brownian motions could be included in \eqref{hestondyn} by considering $W^3:=\rho W^1 + \sqrt{1-\rho^2}W^2$ 
instead of $W^2$.%
~We choose zero correlation 
in order to keep the notation simple.}~The market filtration is the augmentation of the filtration generated by $Y$.%
~The parameters $\mu, \kappa, \lambda,\sigma >0$ will be chosen appropriately later on.%
~We initially assume that the agents are 
trading a single security $S$ in unit endowment-adjusted supply with payoff $S_T=X_T$.%
~We note that, unlike in the original model proposed by Heston, we do not model the log-payoff by \eqref{hestondyn}.
However, our approach is justified by considering only short time horizons.%
~Since the above additive Heston model is affine
and allows for explicit solutions of the functions $\phi$ and $\psi$,
we apply the results obtained in Sections \ref{sec01} and \ref{sec022} to compute the equilibrium price $S_t$ 
at time $t\in[0,T]$ in closed form as a 
function of $Y_t$.
\begin{thm}\label{maintheo}
Let $\theta(\gamma)$ be defined by
\begin{equation*}
\theta(\gamma) =  \left\{\begin{array}{lc}\sqrt{\lambda^2 - \sigma^2\,\gamma^2} &\mbox{if}\e\, \gamma <\frac{\lambda}{\sigma} \\
i \sqrt{ \sigma^2\,\gamma^2 - \lambda^2 } &\mbox{if}\e\, \gamma >\frac{\lambda}{\sigma}\end{array}\right. .
\end{equation*}
Suppose that $\gamma$ is such that $T$ satisfies 
\begin{equation}
T< \left\{\begin{array}{lc}+\infty & \gamma < \frac{\lambda}{\sigma} \\
\frac{2}{|\theta(\gamma)|}\left( \arctan\frac{|\theta(\gamma)|}{-\lambda}  + \pi\right) & \gamma > 
\frac{\lambda}{\sigma}\end{array}\right.\,.\label{maintheo2}
\end{equation} %where $\theta(\gamma)= \sqrt{\sigma^2\gamma^2-\lambda^2}$.
Then we have that, with $\tau:=T-t$, $\theta:=\theta(\gamma)$ and $\theta' := \frac{\partial}{\partial \gamma}\theta(\gamma)$, the equilibrium 
price process $S$ is given by
\begin{equation}\label{maintheo1}
%S_t = T(\tau,\gamma) - \gamma \frac{N(\tau,\gamma)}{D(\tau,\gamma)} \,V_t + X_t \,,
S_t = T(\tau,\gamma) - \gamma \varGamma(\tau,\gamma) \,V_t + X_t \,,
\end{equation}
for $t\in[0,T]$, and where
\begin{multline*}
T(\tau,\gamma) \,= \frac{2\kappa}{\sigma^2\theta }\bigg[\theta(\ef{\theta \tau}+1)+ \lambda(\ef{\theta \tau}-1)\bigg]^{-1}
\left[  \Big(\theta(\ef{\theta \tau}+1)+ \lambda(\ef{\theta \tau}-1)\Big)\Big(\theta' - 
\frac{1}{2}\sigma^2\gamma\tau\Big)\right.\\
- \theta \Big( \theta'(\ef{\theta \tau}+1) + \tau \ef{\theta \tau}(\lambda\theta' - \gamma\sigma^2) \Big) \bigg]\,,
\end{multline*}
\begin{multline*}
\varGamma(\tau,\gamma) = \bigg[\theta \brak{\ef{\theta \tau}+1}+ \lambda\brak{\ef{\theta \tau}-1}\bigg]^{-1} \bigg[
\Big( 2\brak{\ef{\theta \tau}-1}-\gamma\tau\theta' \ef{\theta \tau} \Big) \\
+  \gamma \brak{\ef{\theta \tau}-1} \Big( \theta' \brak{\ef{\theta \tau}+1} + \tau \ef{\theta \tau}\brak{\lambda\theta' 
+ \gamma\sigma^2} \Big)  
\left.\Big(\theta \brak{\ef{\theta \tau}+1}+ \lambda\brak{\ef{\theta \tau}-1}\Big)^{-1}  \right] \,.
\end{multline*}
\end{thm}
We note that \eqref{maintheo2} ensures that \eqref{MAINTHM4} in Theorem \ref{MAINTHM} is satisfied, which, in combination with the
discussion in Section \ref{sec030}, allows us to study the impact of the model parameters in a framework comprising
European-style options.%
~In particular, we illustrate within the Heston framework the effect of the parameters $\gamma$ and $\sigma$ on implied volatilities using
the formulae obtained in Theorem \ref{ThmEx1}.% 
~To this end, we consider a setting
with one underlying asset and fifteen simultaneously traded call options written on it, all affecting the pricing density.%
\begin{figure}[h]
%\centerline{\includegraphics[height=6.2cm]{ImpliVol2}}
\centerline{\includegraphics[width=8.4cm]{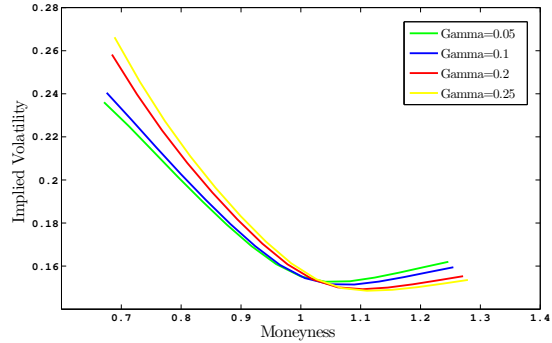}}
\caption{Implied volatility curves with varying risk aversion $\gamma$}\label{fig1}
\end{figure}
\begin{figure}[h]
%\centerline{\includegraphics[height=6.5cm]{VolVol}}
\centerline{\includegraphics[width=8.4cm]{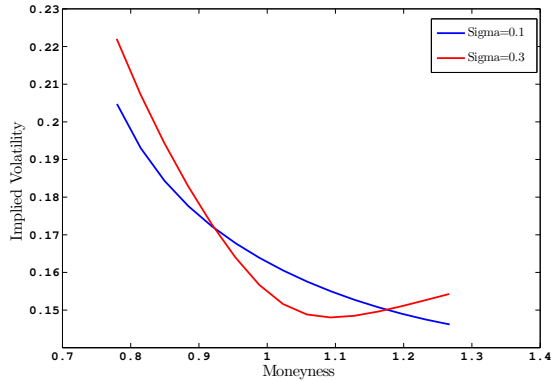}}
\caption{Implied volatility curves with varying vol-of-vol $\sigma$}\label{fig2}
\end{figure}
In Figure \ref{fig1}, four different implied volatility curves are shown, corresponding to
four different values of the risk aversion $\gamma$.
We see that, especially for in-the-money options, higher risk aversion yields
a higher level of implied volatility.%
~The more risk-averse the representative agent is, the more in-the-money options are appreciated 
as good hedges against possibly low values of the underlying.%
~In the recent work \cite{SircarSturm} the impact of market risk aversion on put option implied
volatilities is investigated by means of indifference pricing by dynamic convex risk measures
and asymptotic methods.\\
\indent The implied volatility curves for two different choices of the vol-of-vol parameter $\sigma$ in 
\eqref{hestondyn} are shown in Figure \ref{fig2}.
We observe a significant increase in implied volatility when changing from the low value (blue curve)
to the higher one (red curve).
That is due to the fact that a high value of $\sigma$ increases the probability of $S_T$ taking 
on extreme tail values and hence rendering even out-of-the-money options attractive instruments.\footnote{
For the Figures \ref{fig1} and \ref{fig2}, the following parameters were used for the numerical computations:
$\mu=0.1$, $\kappa=0.006$, $\lambda=0.2$, $T=0.5$, $t=0$, $(x_0,v_0)=(1,0.03)$.
In Figure \ref{fig1}, we set $\sigma=0.3$, whereas in Figure \ref{fig2}, $\gamma=0.2$ was used.}

%In the referenced article, the authors attempt to quantify the
%influence of market participants’ risk preferences by calculating utility in-
%difference prices for options and analyzing the resulting implied volatilities
%through asymptotic methods

%~An increase in the buyers risk aversion results in a decreased implied volatility skew there.
%However, the setting discussed in \cite{SircarSturm} is different to ours in the sense that %setting the option is not traded but rather hedged against
%we consider call options traded in positive supply.

%% file: chap32.tex
\subsection{Equilibrium Dynamics in a Pure Jump Ornstein-Uhlenbeck Setting}\label{sec032}
\begin{figure}[h]
%\centerline{\includegraphics[height=6.2cm]{Jumps2}}
\centerline{\includegraphics[width=8.4cm]{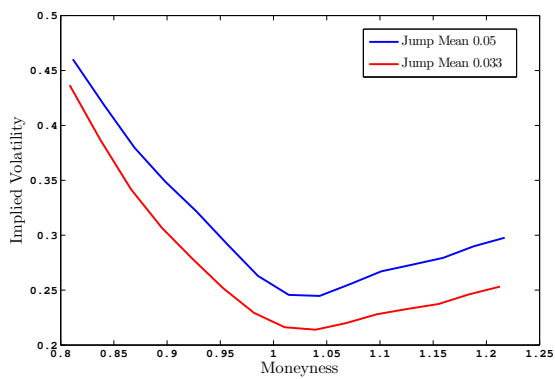}}
\caption{Implied volatility curves with varying jump mean $1/\theta$ and intensity $\kappa$}\label{fig3}
\end{figure}

In order to include the presence of jumps into the discussion of equilibrium prices,
we consider now a single stock with terminal payoff $S_T=X_T$ where %$\F_t=\sigma(X_s,s\le t)$, augmented, for all $t\in[0,T]$, 
%
%where the observable $X$ generating the market filtration is a one-dimensional analytic affine process.% with state space $\re_{\ge 0}$.
%~A classical example,
$X$ is an Ornstein-Uhlenbeck process with a pure jump component as L\'evy 
part\footnote{This is a specific subclass of basic affine processes, compare \citep[Section A.2]{DuffieSingleton01}.}:% or \cite{MKRyield}.}: 
\begin{equation*}%\label{eq321}
dX_t = -\lambda(X_t - \mu)dt + dJ_t\q,\q X_0=x_0\,.
\end{equation*}
Here, $J$ is an adapted compound Poisson process with intensity $\kappa > 0$ and 
jump distribution $\nu(dx) = \frac{1}{2}\theta\exp(-\theta\abs{x})dx$.\footnote{More precisely,
$J_t=\sum_{i=0}^{N_t}b_iD_i$, where $N_t$ is a Poisson process with intensity $\kappa$, $D_i$ are exponentially distributed i.i.d.~random
variables with jumps of mean $\frac{1}{\theta} > 0$, and $b_i$ are i.i.d.~Bernoulli random variables with $P[b_1=1]=P[b_1=-1]=0.5$.}
The parameters $\mu$ and $\lambda$ describe the long term mean and the mean reversion rate, respectively.% 
~In this one-dimensional setting the equations for the functional characteristics $F$ and $R$ 
are given by
\begin{equation}
F(u) = \lambda\mu u + \frac{\kappa\,u^2}{\theta^2 - u^2}  \qq\text{and}\qq  %\\\label{R1}
R(u) = -\lambda u\,,\label{F}
\end{equation}%a
see \eqref{levykin1} and \eqref{levykin2}.%
~Combining \eqref{F} with \eqref{eqapp11} and \eqref{eqapp12}, we deduce that the functions $\phi$ and $\psi$ satisfy 
the following system of Riccati equations
\begin{align*}
\partial_t \phi(t,u)& = \e\lambda\mu \psi(t,u) + \frac{\kappa\,\psi^2(t,u)}{\theta^2 - \psi^2(t,u)}\e,& \phi(0,u)& = 0 \\
\partial_t \psi (t,u)& = \e-\lambda \psi(t,u)\e,&  \psi(0,u)& = u\,\,,
\end{align*}
which allows for the explicit solutions
\begin{equation*}%as
\phi (t,u) = \frac{\kappa}{2\lambda}\log\brak{\frac{\theta^2 - u^2\ef{-2\lambda t}}{\theta^2 - u^2}} + \mu u (1-\ef{-\lambda t}) \qq\text{and}\qq
\psi (t,u) = u\ef{-\lambda t}.
\end{equation*}%as
Thus, \eqref{affinedef2} holds, as long as $u \in\R\backslash\{-\theta,\theta\}$ %\in [-\infty,\theta)$ 
and $T < t^*(u)$, with 
\begin{eqnarray}\label{eq322}
t^*(u) = \left\{\begin{array}{lc} +\infty \q & \abs{u}<\theta \\ -\frac{1}{2\lambda}\log(\frac{\theta^2}{u^2}) \q& \abs{u}>\theta \end{array} \right.\,.                                                                                            
\end{eqnarray}
This, together with Corollary \ref{AffPropCor}, allows us to formulate the following:
\begin{prop}\label{prop321}
If $\abs{{\tg}}\neq \theta$ and $T<t^*(-{\tg})$, where $t^*$ is as in \eqref{eq322}, then, with $\tau:=T-t$,  the equilibrium 
price process $S$ is given by
\begin{equation*}
S_t = \left[    \frac{\kappa\theta^2{\tg}\brak{\ef{-2\lambda \tau} - 1}}{\lambda(\theta^2 - {\tg}^2)\brak{\theta^2-u^2\ef{-2\lambda\tau}}}   
+ \mu (1 - \ef{-\lambda \tau} )\right] + \ef{-\lambda \tau}X_t \,,\q t\in[0,T]\,.
\end{equation*}
\end{prop}
In the following we illustrate the influence of the parameters $\gamma$, $\kappa$ and $\theta$ on option implied volatilities.
Figure \ref{fig3} illustrates the dependence of implied volatilities on the jump parameters for fixed risk aversion.
\begin{figure}[h]
%\sidecaption
%\centerline{\includegraphics[width=8.4cm]{ra_ou20}}\\
\centerline{\includegraphics[width=8.4cm]{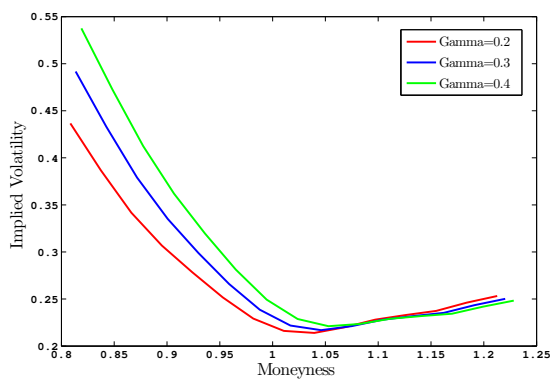}}
\caption{Implied volatility curves with varying risk aversion $\gamma$}\label{fig3a}
\end{figure}
The red curve corresponds to smaller jumps arriving at a high frequency ($(\kappa,\frac{1}{\theta})=(30,\frac{1}{30})$), 
whereas the blue one was obtained considering higher jumps at a lower 
frequency ($(\kappa,\frac{1}{\theta})=(20,\frac{1}{20})$).%
~Increasing the mean jump height distinctly lifts the level of implied volatility, since the probability
of $S_T$ taking extreme values is higher that way.%
~We further note that an affine model including jumps seems in general more suitable to reproduce the
right-hand side smile observed in real market data.
In Figure \ref{fig3a} in turn, we observe that an increase in implied volatility for in-the-money call options is caused by
increasing risk aversion, similar to the stochastic volatility model discussed before.\footnote{The remaining parameters 
in Figures \ref{fig3} and \ref{fig3a} were chosen as $(\mu,\lambda,T,t,x_0)=(1,2,0.1,0,1)$.%
~In Figure \ref{fig3} we set ${\gamma}=0.2$, whereas the jump parameters were chosen as
($\kappa,\frac{1}{\theta})=(30,\frac{1}{30})$) in Figure \ref{fig3a}.~As before, we considered 15 simultaneously traded call options.}
%This effect occurs independently of the ratio of jump size and jump frequency, which was again chosen 
%as ($\kappa,\frac{1}{\theta})=(20,\frac{1}{20})$), 
%and ($\kappa,\frac{1}{\theta})=(30,\frac{1}{30})$) in the upper and lower graphic, respectively.
%Note that the different levels for different jump parameters remain observable.

%% file: chap021.tex
\section{Information-Based Equilibrium Pricing}\label{sec021}
In this section, we propose another method to model the market filtration based on the information-based asset 
pricing approach of \cite{andrea04} and \cite{andrea01}. 
This approach is based on the modeling of cash flows and the explicit construction of
market filtrations, which can be naturally embedded in the equilibrium pricing model considered in 
the present paper. 
The key idea is that, instead of assuming from the outset some abstract filtration representing the information available to the market, 
processes carrying market-relevant information are explicitly constructed,
and a distinction between "genuine" information and market noise is made.
The equilibrium dynamics is then computed by using the special form of the pricing measure obtained in Section
\ref{sec01}, by assuming an a priori distribution of the market factor determining the terminal payoff, and 
by updating a posteriori distributions about the assets' payoffs  obtained by a version of
Bayes formula.

\subsection{Setup and Equilibrium Pricing Formula}

We assume that the
probability space $(\Omega,\F,P)$ supports a $N$-dimensional
Brownian motion $B$
together with $N$ independent random market factors $(X_i)_{i=1}^N$, all independent
of $B$, and define $S^k_T = f^k(X_1,\dots,X_N)$.
The agents know the a priori distributions $\nu^i$ of all $X_i$. 
With each market factor $X_i$, we associate an observable process $(\xi_t)_{t \in [0,T]}$, 
the so-called information process. 
The information processes are defined by
\begin{equation}\label{eq2101}
\xi^i_t = \sigma_iX_i t + \beta^i_t\,,\q t\in[0,T]\,,
\end{equation}
where the independent standard Brownian bridges $\beta^i$ on $[0,T]$ are
defined in terms of $B$ as solutions to
the SDEs
\begin{equation}\label{eq2101a}
d\beta^i_t = - \frac{\beta^i_t}{T-t}dt +dB^i_t\,,\q \beta^i_0=0\,,
\end{equation}
for $t\in[0,T)$, and $\beta^i_T=0$.%
~Looking at the different components of the processes \eqref{eq2101}, we identify the part $\sigma_iX_i t$
containing real information about the realization of a market factor revealed
over time, and the bridge part representing market noise.%
~The speed at which the outcome of $X_i$ is revealed is governed by the information rate $\sigma_i$.
The information processes capture the flow of information available to
the market agents, and thus generate the market filtration:
\begin{equation*}%\label{eq2102}
\F_t = \sigma\brak{\xi^1_s,\dots,\xi^N_s,s\le t}\,,\q t\in[0,T]\,.
\end{equation*}
By construction, $S_T$ is $\F_T$-measurable, and at each time $t\in[0,T]$, the equilibrium price $S_t$ will be determined
using the results of Section \ref{sec01}.
\begin{thm}\label{thm2101}
Assume that all a priori distributions $\nu^i$ allow for a density with respect to the
Lebesgue-measure denoted by $v^i(x)$, respectively.%
~If in addition the functions $(f^k)^K_{k=1}$ and the a priori densities $(v^{i})_{i=1}^N$ are such that \eqref{equiS1} is satisfied, then,
for $t<T$, the equilibrium price process of the $k$-th security is given by
\begin{equation}\label{eq2104}
S^k_t = \frac{\int_{\R^N}\, z(x_1,\dots,x_N)f^k(x_1,\dots,x_N)
\pi^1_t(x_1)\cdots\pi^N_t(x_N)dx_1\ldots dx_N}
{\int_{\R^N} \,z(x_1,\dots,x_N)\pi^1_t(x_1)\cdots\pi^N_t(x_N)dx_1\ldots
dx_N}\,,
\end{equation}
where the function $z$ is defined by
\begin{equation}\label{eq2105}
z(\cdot) = \exp\edg{-\sum_{l=1}^K\tg^lf^l(\cdot)}.
\end{equation} 
The regular conditional density function $\pi^i_t$ associated with the $i$-th
market factor is given by
\begin{equation}\label{eq2106}
\pi^i_t(x) = \frac{v^i(x)\exp\edg{\frac{T}{T-t}\brak{\sigma_i x \xi^i_t -
\frac{1}{2}(\sigma_i x)^2 t}}}
{\int_\R v^i(y)\exp\edg{\frac{T}{T-t}\brak{\sigma_i y \xi^i_t -
\frac{1}{2}(\sigma_i y)^2 t}}dy}\,.
\end{equation}
\end{thm}

\subsection{Innovation Processes and Equilibrium Market Price of Risk}
Let us consider $K=N=1$, and in particular the case $S_T=X$
with corresponding information process $\xi_t = \sigma X t + \beta_t$, for $t\in[0,T]$.%
~We assume that the market factor $X$ is such that the conditions of Theorem \ref{thm2101} are satisfied. Formula \eqref{eq2104} now reduces to
\begin{equation}\label{eq2104a}
S_t= \frac{\EF{S_T \exp\brak{-\tg S_T}}{t}}{\EF{\exp\brak{-\tg S_T}}{t}} = \frac{\int x \exp\brak{-\tg
x} \pi_t(x)dx}{\int \exp\brak{-\tg x}\pi_t(x)dx}\,.
\end{equation}
Results from general filtering theory guarantee the existence of a
$P$-Brownian motion $W$ on $[0,T)$, adapted to
the market filtration generated by $\xi$.%
~Observe to this end that rearranging \eqref{eq2101a} leads to the following SDE
satisfied by $\xi$ on $[0,T)$
\begin{equation*}%\label{eq2107}
d\xi_t = \edg{\frac{1}{T-t}\brak{\sigma T X - \xi_t}}dt + dB_t\,,\q\xi_0=0\,.
\end{equation*}
Hence, $W$ is the innovations process associated with the information
$\xi$ given by
\begin{equation}\label{eq2108}
W_t= \xi_t - \int_{0}^{t}{\edg{\frac{1}{T-s}\brak{\sigma T \EF{X}{s} - \xi_s}}ds}\,,\q t<T\,.
\end{equation}
Thus, instead of having to assume the existence of Brownian motions as drivers for the prices, they rather emerge
naturally from within the information-driven structure, as the following proposition shows.
\begin{prop}\label{prop3201}
Assume that $g(X)$ and $h(X)$ belong to $L^2(P)$ where $g(x) = x\exp(-\tg x)$ and $h(x) = \exp(-\tg x)$.
Then the equilibrium dynamics of $(S_t)_{t<T}$ are given by
%\begin{equation}\label{eq2111}
%dS_t = \sigma^h_t\edg{\sigma^h_t\brak{S_t(V^h_t)^2 - V^g_tV^h_t}dt +
%\brak{V^g_t-S_tV^h_t}dW_t}\,,
%\end{equation}
\begin{equation}\label{eq2111}
dS_t = \frac{\sigma T}{T-t}{\rm Var}^Q_{t}(X)\edg{\frac{\sigma T}{T-t}\brak{\EF{X}{t} - S_t}dt + dW_t}
\end{equation}
where
\begin{equation}\label{eq2110}
{\rm Var}^Q_{t}(X) := \EFQ{X^2}{t} - \brak{\EFQ{X}{t}}^2
\end{equation}
is the conditional variance of $X$ under the measure $Q$ defined in \eqref{density2}.
%\begin{equation*}
%\sigma^h_t=\frac{\sigma T}{(T-t)\EF{h(X)}{t}}\,.
%\end{equation*}
%For a function $\varphi:\R\rightarrow\R$ such that $\varphi(X) \in L^2(P)$,
%\begin{equation}\label{eq2110} 
%V^\varphi_t = \EF{\varphi(X)X}{t} - \EF{\varphi(X)}{t}\EF{X}{t}
%\end{equation}
%is the conditional covariance of the market factor with the function $\varphi$.
\end{prop}
%The expressions $V^g_t,V^h_t$ and $\EF{h(X)}{t}$ can be worked out
%semi-explicitly by means of \eqref{eq2110},
%the integral formula \eqref{eq2104a}, and the regular conditional density
%$\pi(x)$ defined in \eqref{eq2106}.
%Each of them is a function of $t$ and $\xi_t$ due to the Markov property of the
%information process.
The expressions $E[X|\F_t]$ and ${\rm Var}^Q_{t}(X)$ can be worked out
semi-explicitly by means of \eqref{eq2110},
the integral formula \eqref{eq2104a}, and the regular conditional density
$\pi(x)$ defined in \eqref{eq2106}.
They are functions of the pair $(t,\xi_t)$ and triplet
$(t,\xi_t,\tg)$, respectively, due to \eqref{density2} and the Markov property of the
information process.
By an application of L\'evy's characterization of Brownian motion, it can be shown that the process $(W^Q_t)_{t<T}$ defined by
\begin{equation*}
dW^Q_t = \frac{\sigma T}{T-t} \brak{\EF{X}{t} - S_t} dt +dW_t 
\end{equation*}
is an $((\F_t),Q)$-Brownian motion.~Thus, \eqref{eq2111} confirms that $(S_t)_{t<T}$ is an $((\F_t),Q)$-martingale.

\subsection{Pricing Credit-Risky Securities}\label{sec0213}
In this section, we illustrate the impact of the ``noisyness'' of information and of
the market risk aversion on the equilibrium prices of 
a credit-sensitive security within a simple benchmark model, see \cite{andreaHAZARD}, where the a-priori distribution of $S_T=X$ is discrete: 
$S_T \in \{x_0,x_1\}=\{0,1\}$. We denote by  $p_0 := P[X=0]$ the probability of default.%
~Due to the discrete payoff structure, formula \eqref{eq2104} simplifies and
allows us to examine the impact of model parameters,
such as the information flow rate or the risk aversion, on the equilibrium price of $S$.%
~The price of the security threatened by default can be obtained in closed form analogously to
\eqref{eq2104} and is given by
\begin{equation*}%\label{eq2112}
S_t = \frac{ p_1 x_1 \exp\brak{-\tg x_1} \exp\edg{\frac{T}{T-t}\brak{\sigma x_1 \xi_t -
\frac{1}{2}(\sigma x_1)^2 t}}}
{\sum_{i=0,1} p_i \exp\brak{-\tg x_i} \exp\edg{\frac{T}{T-t}\brak{\sigma x_i \xi_t -
\frac{1}{2}(\sigma x_i)^2 t}}}\,,\q t<T\,.
\end{equation*}
%Applying It\^o's product rule to \eqref{eq2112} together with
%\eqref{eq2108} yields
%the following dynamics%\footnote{The expression ${\rm Var}_{Q,t}(X)$ denotes the conditional variance of $X$ under $Q$, given $\F_t$.} 
%for $S$
%\begin{equation*}%\label{eq2113}
%dS_t = \frac{\sigma T}{T-t}{\rm Var}_{Q,t}(X)\edg{\frac{\sigma
%T}{T-t}\brak{\EF{X}{t} - S_t}dt + dW_t}\,,\q t<T\,.
%\end{equation*}
%Due to the Markov property of the information process, the terms $\EF{X}{t}$
%and ${\rm Var}_{Q,t}(X)$ 
%are functions of the pair $(t,\xi_t)$ and triplet
%$(t,\xi_t,\tg)$, respectively.

\begin{figure}[h]
%\sidecaption
\centerline{\includegraphics[width=8.4cm]{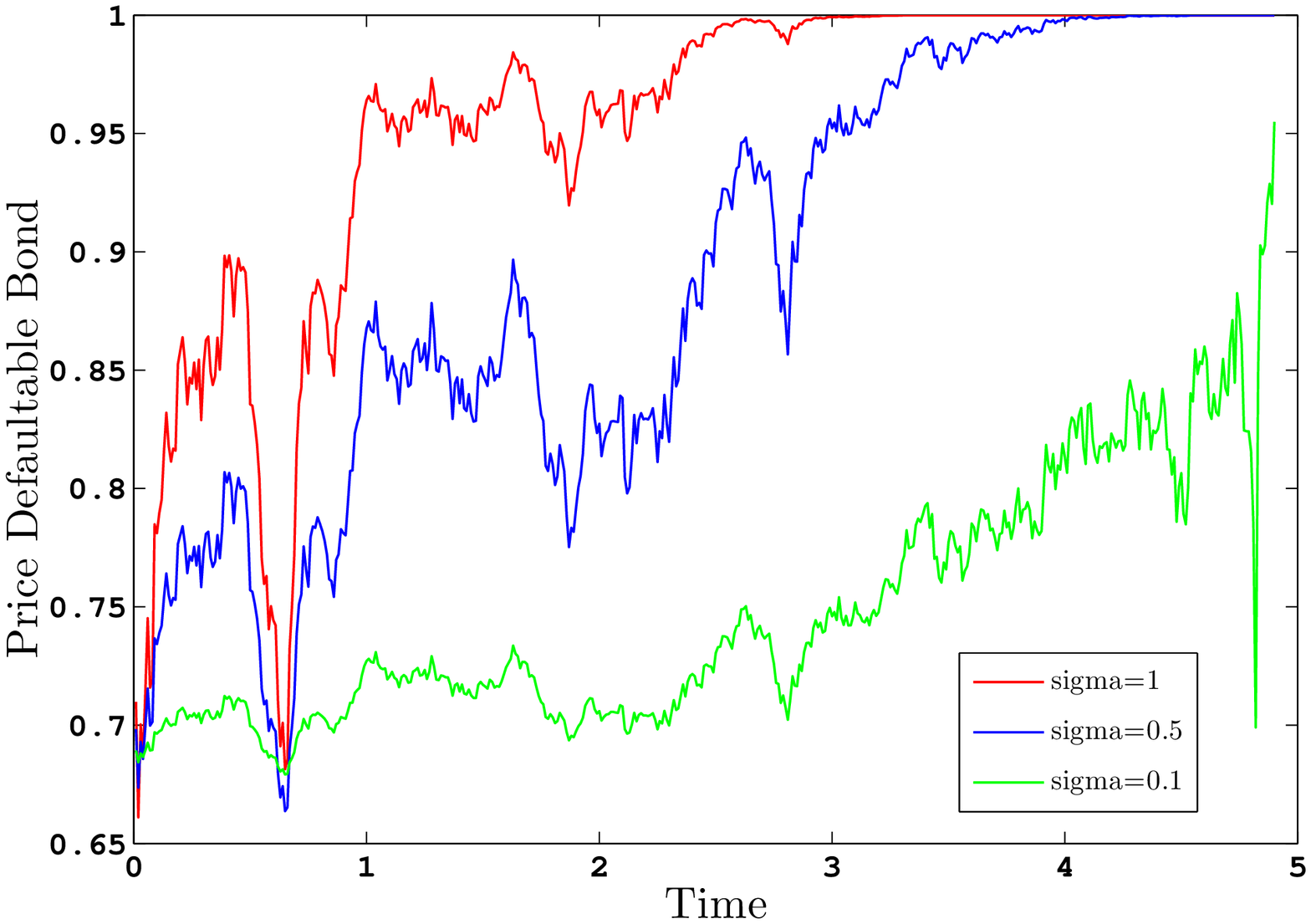} \includegraphics[width=8.4cm]{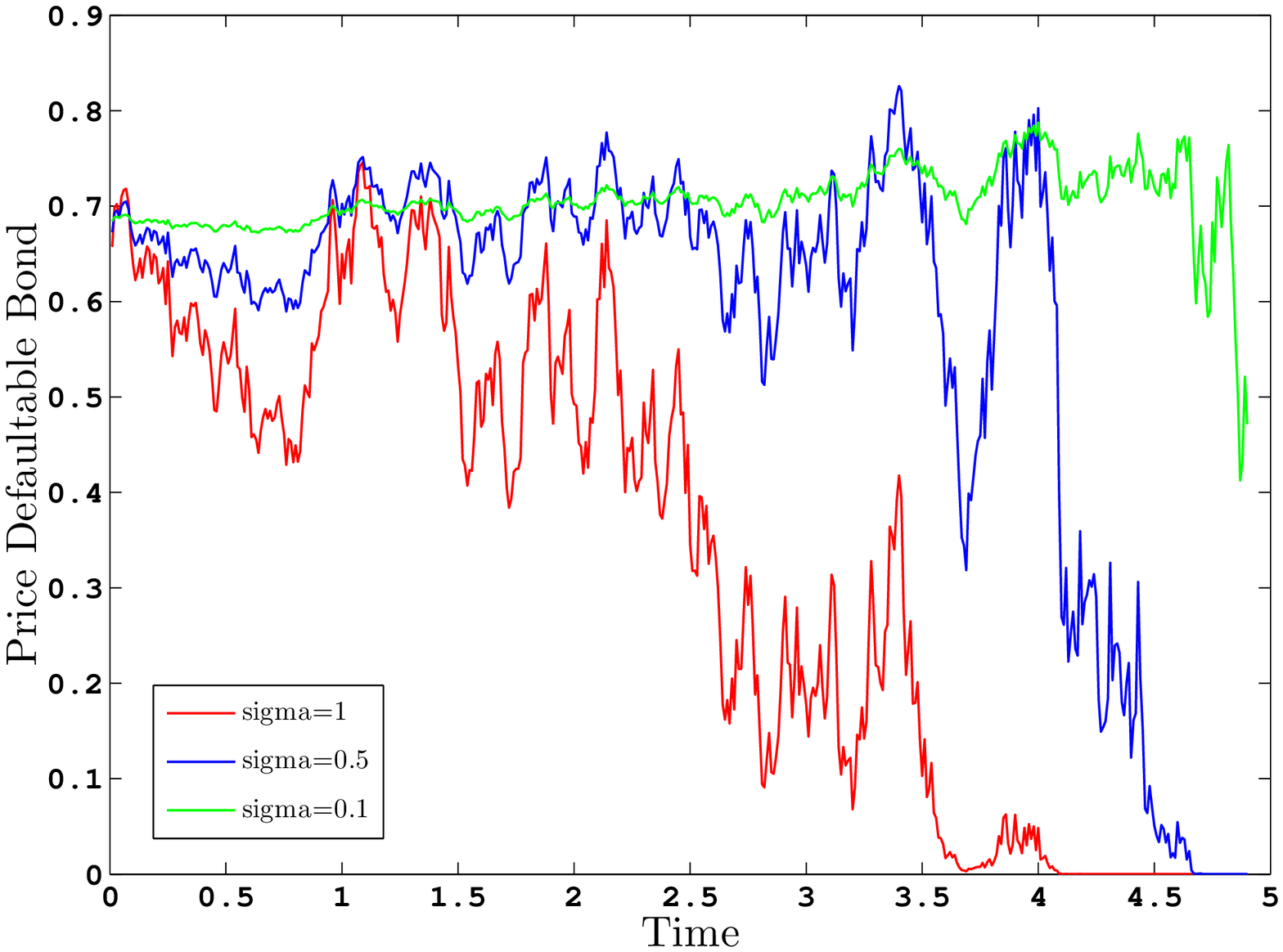}}
\caption{Defaultable bond prices: influence information rate $\sigma$}\label{fig4}
\end{figure}

Figure \ref{fig4} shows the impact of $\sigma$ on the price of a defaultable bond, where the
probability of default is chosen to be $p_0=0.2$.
\begin{figure}[h]
%\sidecaption
\centerline{\includegraphics[width=8.4cm]{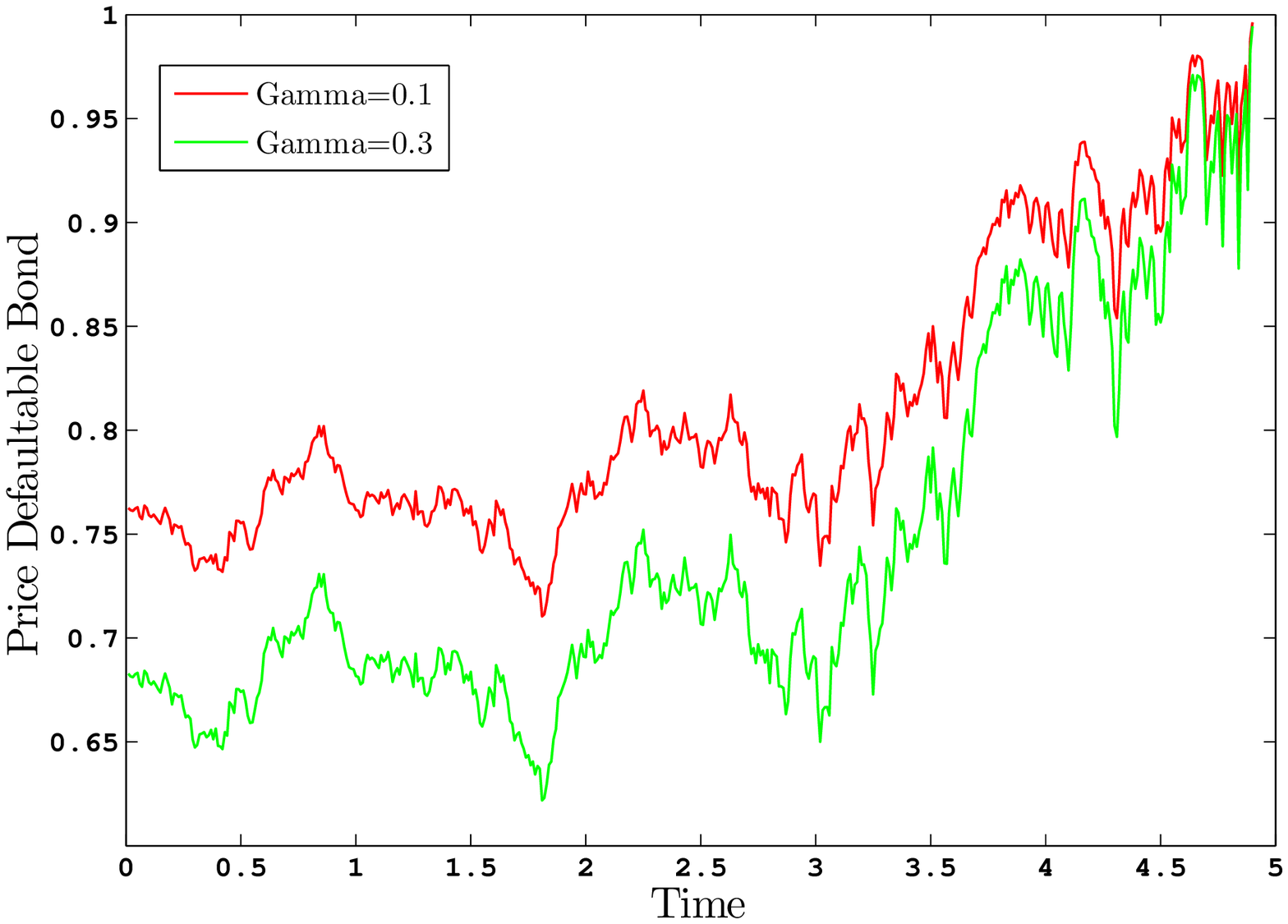} \includegraphics[width=8.4cm]{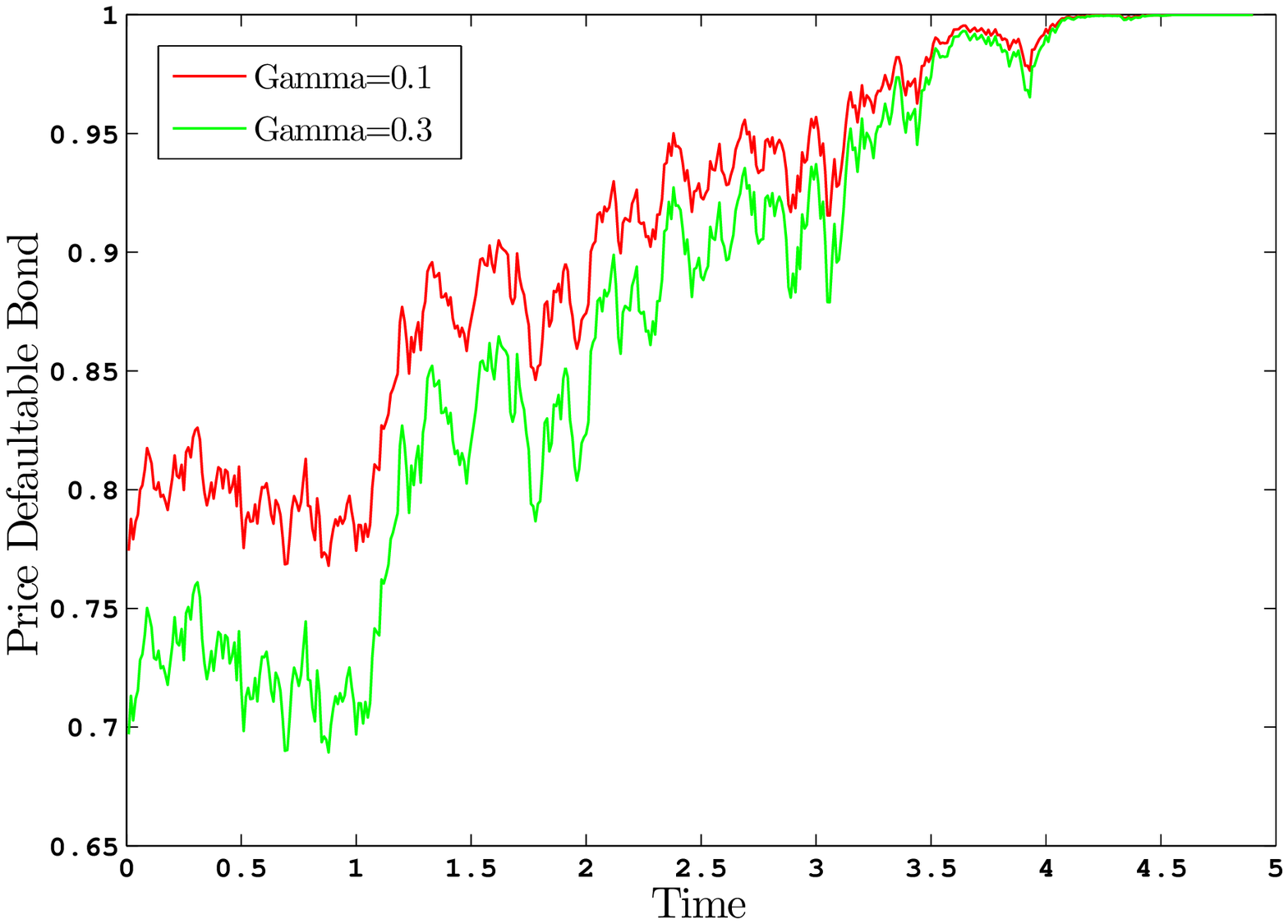}}
\caption{Defaultable bond prices: influence risk aversion $\gamma$}\label{fig5}
\end{figure}
In the upper graphic the bond does not default, whereas in the lower graphic we considered
the situation of a default.
In both cases, a low information flow rate (green curve, $\sigma=0.1$) leads to a 
rather late adjustment of the equilibrium price process towards the prevailing terminal value, while 
the red curve ($\sigma = 1$) reacts earlier to the information about the outcome of $X$.%
~The influence of the risk aversion $\tg$ on defaultable bond prices is demonstrated in Figure \ref{fig5}. 
It is evident that a higher risk aversion leads to a more careful evaluation of the bond, since the possibility
of a default is taken more into account.
This effect occurs in both depicted scenarios, where in the upper and lower figure the information rate $\sigma$ 
is chosen to be $\sigma=0.2$ and $\sigma=0.5$, respectively.
Note however that for the case of a low information rate (upper graphic) the initial price difference
turns out to be smaller, because both agents, the more and less risk-averse one, 
consider the information to be noisier, hence less valuable, and thus give the bond a lower price.\footnote{The 
following parameters were used for the simulations shown in Figures \ref{fig4} and \ref{fig5}:~$P[X=x_1]=0.8$, $T=5$.
The price process is shown for $t\in[0,4.9]$.~In Figure \ref{fig4} we set $\tg=0.6$.}

\subsection{One-Dimensional, Exponentially-Distributed Terminal Cash Flow}\label{example211}
We illustrate how, for particular choices of $v$ and $f$, the formulae
\eqref{eq2106} and \eqref{eq2104a} 
can be worked out explicitly.
We assume $f(x) = x$, corresponding to the assets payoff itself being the market
factor.% 
~Furthermore, the a priori distribution of $S_T$, the cash flow at time $T$, is assumed to be exponential.
\begin{cor}\label{cor2101}
Assume that the a-priori distribution of $S_T=X$ is of the exponential form,
that is, $v(x) = (1_{\{x\ge0\}}/\kappa)\exp\brak{-x/\kappa}$ 
for some $\kappa > 0$.%
~If $\tg > \kappa - 1$,
then the equilibrium price at time $t<T$ is given by
\begin{equation}\label{eq2114}
S_t = \edg{\frac{\exp\brak{-\frac{1}{2}B_t^2/A_t}}{\sqrt{2\pi A_t}
\,\mathcal N(B_t/A_t)} + \frac{B_t}{A_t}}\,\,,
\end{equation}
where 
\begin{equation}\label{eq2115}
A_t = \sigma^2tT /(T-t) \q,\q B_t = \sigma T \xi_t/(T-t) -
\frac{\tg\kappa + 1}{\kappa}\,,
\end{equation}
and $\mathcal N(x)$ denotes the standard normal distribution function.
\end{cor}
Since the pricing measure depends only on the terminal cash-flow as a consequence
of the attainable endowments, changing from 
$P$ to $Q$ could be interpreted as a different view $\tilde v$ of the
representative agent
on the a-priori-distribution of $S_T$.
More precisely, under $Q$ the cash-flow $S_T$ is exponentially distributed with
new parameter $(\tg \kappa +
1)/\kappa$, also appearing in \eqref{eq2115}, which can be seen by
working out the adjusted density 
\begin{equation}\label{eq2116}
\tilde v (x) = \frac{\exp(-\tg
x)v(x)}{\int \exp(-\tg y)v(y)dy}\,.
\end{equation}

%% file: appendix01.tex
\section{Proofs}\label{appendix01}
{\bf Proof of Theorem \ref{density}}\\
\noindent Due to the time-consistency and strict monotonicity of the entropic preferences, it suffices to
show that the strategies $\hv^a$ are optimal for the utility maximization in $t=0$.%
~Note first that \eqref{equiS1} ensures that \eqref{density2} and \eqref{equiS} are 
well-defined.%
~In particular, the price process $S$ is a  $Q$-martingale, and thus $Q\in\mathcal P$.%
~Furthermore, the constant strategies $\hv^a$ lie in $\Theta$, since for any $\tilde Q\in \mathcal P$, the process
$G_t(\hv^a)=\hv^a\cdot(S_t-S_0)$ is by assumption a $\tilde Q$-martingale, 
and hence in particular a $\tilde Q$-supermartingale.
\\[5pt]
We now show that the quantity $\gamma$ introduced in \eqref{revi02} can be seen as the risk 
aversion of some representative agent whose optimal utility is attained at the
constant strategy $\vt^*\equiv n + \eta$.%
~Indeed, since $S$ is a $Q$-martingale, and $(n+\eta)\cdot S_T\in L^1(Q)$,
the utility maximization of the representative agent can be formulated as follows\footnote{Note that the first expression in \eqref{p4} is
equivalent to the representative agent's utility maximization
of terminal wealth against both, the aggregated initial endowments $\eta$ and aggregated net supply $n$, over all admissible strategies.}:
\begin{align}
&\e\sup_{\vt\in\Theta, \EQ{G_T(\vt)}\le\EQ{(n+\eta)\cdot S_T}} \crl{ U_0^\gamma\big(G_T(\vt)\big) }\nonumber\\
\le&\qq\qq\q\,\,\, \sup_{\vt\in\Theta} \crl{ U_0^\gamma\big(G_T(\vt) - \EQ{G_T(\vt)} + \EQ{(n+\eta)\cdot S_T}\big)} \nonumber\\
=&\qq\qq\q\,\,\, \sup_{\vt\in\Theta} \crl{ U_0^\gamma\big(G_T(\vt)\big) - \EQ{G_T(\vt)}}  + \EQ{(n+\eta)\cdot S_T}\nonumber\\
\le&\qq\qq\q\,\,\,  \frac{1}{\gamma}H(Q|P) + \EQ{(n+\eta)\cdot S_T}\,.\label{p4}
\end{align}
The last inequality is derived from the dual representation of $U_0^\gamma$, where the relative entropy
is given by $H(Q|P)=E[\frac{dQ}{dP}\log(\frac{dQ}{dP})]$.%
~But $G_T(\vt^*)$ with $\vt^*\equiv n+\eta$ plugged into the representative agent's 
utility $U^\gamma_0(\cdot)$ yields
\begin{equation*}
U^\gamma_0\big((n+\eta)\cdot S_T\big) 
= \frac{1}{\gamma}H(Q|P) + \EQ{(n+\eta)\cdot S_T} \,.
\end{equation*}
Comparing this with \eqref{p4} shows that $\vt^*\equiv n+\eta$
is indeed optimal for the representative agent when the price process $S$ is given by \eqref{equiS}.%
~Individual optimality of $\hv^a$ for the single agents now follows by a scaling argument
and the specific form of the aggregated endowment.%
~Note that, for all $a\in\mathbb A$,
\begin{equation*}
\e\e \vt^*=\argmax_{\vt\in\Theta}\crl{U_0^\gamma\big(G_T(\vt)\big)}
\end{equation*}
is equivalent to
\begin{equation*}
\frac{\gamma}{\gamma^a} \vt^*=\argmax_{\vt\in\Theta}\crl{U_0^a\big(G_T(\vt)\big)}\,,
\end{equation*}
which in turn is equivalent to
\begin{equation*}%\label{p1}
\frac{\gamma}{\gamma^a} \vt^* - \eta^a = \argmax_{\vt\in\Theta}\crl{U_0^a\big(H^a + G_T(\vt)\big)}\,.
\end{equation*}
This shows that $\hv^a$ is the optimal strategy for agent $a \in {\mathbb A}$.%
~Since the
strategies $(\hv^a)_{a\in\mathbb A}$ add up to $n$, the market clears at any time, and
hence the pair $((S_t)_{t\in[0,T]},(\hv^a)_{a\in\mathbb A})$ forms an equilibrium.\hfill\ensuremath{\square}\\

\noindent{\bf Proof of Theorem \ref{MAINTHM}}\\
\noindent  \emph{Part 1: Pricing Formula \eqref{mainTHMform}.}
From Section \ref{ap} it is known that $Y=(V,X)$ satisfies
\begin{equation}\label{affinetrans}
\EF{\exp\brak{u\cdot Y_T}}{t} = \exp\edg{\phi(\tau,u) + \psi(\tau,u)\cdot Y_t}\,,%\tag{A.2}
\end{equation}%nd{align}
for all $u=(u_v,u_x) \in \C^2$ such that $(T,u) \in \mE_{\C}$, since the latter implies that \eqref{affinedef1} and thus \eqref{affinedef2}
hold for all $t\in[0,T]$.
\ak{
By assumption, %and with $V:=(V^1,\dots,V^{m})$, 
the process $Y=(V,X)$ is analytic affine and hence we know from Section \ref{sec022}
that its conditional characteristic 
function allows for the representation 
\begin{equation*}\label{affinetrans}
\EF{\exp\brak{u\cdot Y_T}}{t} = \exp\edg{\phi(\tau,u) + \psi(\tau,u)\cdot Y_t}\e,\tag{A.2}
\end{equation*}%nd{align}
for all $u=(u_v,u_x) \in \C^2$ such that $(T,u) \in \mE_{\C}$\,.%
~This holds, since the fact that $\D_t \supseteq \D_T$, whenever $t\le T$, and $(T,u) \in \mE_{\C}$ imply that formula \eqref{affinedef1} holds for
$t$, whenever it holds for $T$, and hence \eqref{affinedef2} as well.
}
\\[5pt]
Let us assume for the moment that \eqref{equiS1} holds.%and that $S_T\in L^1(Q)^K$.
~This will be verified later.~We 
then know from \eqref{density2} that the equilibrium pricing measure $Q$ is given by its Radon-Nikodym-density
\begin{align*}%\label{App21}
\frac{dQ}{dP} = \frac{\exp\brak{-\tg\cdot S_T}}{\E{\exp\brak{-\tg\cdot S_T}}} 
= \frac{\exp\brak{-\tg\cdot f(X_T)}}{\E{\exp\brak{-\tg\cdot f(X_T)}}}\,.
\end{align*} 
Hence, by applying Bayes formula and following \eqref{equiS}, we obtain
\begin{equation}\label{App22}
S^k_t = \EFQ{ S^k_T}{t} =  \frac{\EF{f^k(X_T) \exp\brak{-\tg\cdot f(X_T)}}{t}}
{\EF{\exp\brak{-\tg\cdot f(X_T)}}{t}}%\tag{A.3}
\end{equation}%nd{align}
for the equilibrium price of the $k$-th security.
The Fourier transforms $\hat g^k$ and $\hat h$ defined in  \eqref{MAINTHM2} and \eqref{MAINTHM3}, respectively, exist 
and are integrable by assumption.%\footnote{Compare e.g. REFERENCE.}
~Hence we apply the Fourier inversion formula\footnote{See \citep[Theorem 9.5.4]{dudley01}.} to obtain
\begin{equation*}
g^k(x) = \frac{1}{2\pi} \int_{\R} \ef{isx} \hat g^k(s) ds\e\q\mbox{and}\e\q 
h(x) = \frac{1}{2\pi} \int_{\R} \ef{isx} \hat h(s) ds \,,
\end{equation*}%nd{align*}
$dx$-almost surely. With this at hand, \eqref{App22} transforms to 
\begin{equation}
S^k_t = \frac{\EF{\exp\brak{-\alpha^k X_T} g^k(X_T)}{t}}{\EF{\exp\brak{-\beta X_T} h(X_T)}{t}} 
= \frac{\EF{ \int_{\R} \exp\edg{(-\alpha^k + is)X_T} 
\hat g^k(s) ds}{t}}{\EF{ \int_{\R} \exp\edg{(-\beta + is)X_T} \hat h(s) ds}{t}} \,.\label{App23}
\end{equation}
Now we observe that
\begin{multline}\label{ig}
\EF{ \,\,\bigg| \,\int_{\R} \exp\edg{(-\alpha^k + is)X_T} \hat g^k(s) ds \,\,\bigg| \,\,}{t} \\
< \EF{ \exp\brak{-\alpha^k X_T} \int_{\R} \abs{\hat g^k(s)}  ds \,}{t} < \infty\,,\end{multline}
since $\brak{T,(0,-\alpha^k)}\in \mE\subseteq\mE_{\C}$ and $\hat g^k$ is integrable.%
~The same holds analogously for the denominator in \eqref{App23}.%
~In particular, we have 
\begin{equation}\label{revi1}
0< \EF{\exp\brak{-\tg\cdot f(X_T)}}{t} < \infty\,,\e \mbox{for all} \e t\in[0,T]\,,%\tag{A.6}
\end{equation}
since we required $Y^T$ to be conservative and $\brak{T,(0,-\beta)}$ to lie in $\mE$.%
~Thus, \eqref{ig} and \eqref{revi1}, in combination with \eqref{App22}, yield that \eqref{equiS1} is indeed satisfied.%
~We may now apply Fubini`s Theorem to exchange the order of integration, and we get that
%\begin{multline}
\begin{multline}
\EF{ \int_\R \exp\edg{(-\alpha^k + is)X_T} \hat g^k(s) ds}{t} =  \int_\R \EF{ \exp\edg{(-\alpha^k + is)X_T}}{t} \hat g^k(s) ds\\
= \int_\R \exp\edg{\phi\big(\tau,(0,-\alpha^k+is)\big) 
+\psi\big(\tau,(0,-\alpha^k+is)\big)\cdot Y_t} \hat g^k(s) ds\,\,.\label{App24}
\end{multline}
%\end{multline}
The affine transformation formula \eqref{affinetrans} holds, since $\brak{T,(0,-\alpha^k)} \in \mE$.~Applying
the same arguments to the denominator in \eqref{App23} combined with \eqref{App24} yields the desired form of $S^k_t$
in \eqref{mainTHMform}.\\

\noindent \emph{Part 2: Pricing Formula \eqref{AffPropEQ}.}
~We outline the details for $K=1$, the rest follows
by repeating the arguments for the partial derivative with respect to each $\zeta^k$.%` evaluated at $\tg$.
~So we assume we only have one security
$S$ with corresponding $\tg \in \R$ affecting the density of the pricing measure $Q$.%
~It follows that
\begin{equation*}
\frac{dQ}{dP} = \frac{\exp(-\tilde\gamma S_T)}{\E{\exp(-\tilde\gamma S_T)}} 
= \frac{\exp(-\tilde\gamma f(X_T))}{\E{\exp(-\tilde\gamma f(X_T))}}\end{equation*}
and the equilibrium price of $S$ at time $t$ can be obtained again by computing 
\begin{equation}\label{AffProp2}
S_t = \frac{\EF{f(X_T) \exp(-\tilde\gamma f(X_T))}{t}}
{\EF{\exp(-\tilde\gamma f(X_T))}{t}}\,.%\tag{A.8}
\end{equation}
Recall from Part 1 that
\begin{equation}\label{AffProp4}
\exp\brak{-\tg f(X_T)} \in L^1(P)  \q\text{and}\q f(X_T)\exp\brak{-\tg f(X_T)} \in L^1(P)\,,%\tag{A.9}
\end{equation}
due to the assumption of $(T,(0,-\alpha))$ and $(T,(0,-\beta))$ lying in $\mE$.%
~Since the set $\mE_{\C}$ is open, compare  
\citep[Lemmata 3.12 and 3.19]{MKRphd}, and due to the integrability assumptions
on the functions $g^k_\zeta(s)$, the first integrability in \eqref{AffProp4} even holds in some neighbourhood of $\tg$,
allowing us to differentiate the function $\zeta \mapsto E[\exp(-\zeta f(X_T))|\F_t]$ at $\zeta=\tg$.%\footnote{Here, we may assume without loss 
%of generality that the model dependent "damping factor" $\beta$ is chosen according to a mapping $\tg\mapsto\beta(\tg)$ that
%is differentiable in a sufficiently small neighbourhood of $\tg$.}%
~Indeed,
by the smoothness of the mapping $\zeta \mapsto \exp(-\zeta f(X_T))$ and the integrability of the second term in \eqref{AffProp4}, we obtain 
\begin{equation}\label{AffProp6}
\EF{ f(X_T) \exp\brak{-\tg f(X_T)} }{t}  = -\left.\frac{\partial}{\partial \zeta}\EF{ \exp\brak{-\zeta f(X_T)}}{t}\right|_{\zeta=\tg}\,,%\tag{A.10}
\end{equation}
as an application of the triangular inequality and dominated convergence.
On the other hand we know from an analogue of \eqref{App23} and \eqref{App24} that the 
denominator in \eqref{AffProp2} can be computed by  %Theorem \ref{MAINTHM} that
\begin{multline}\label{AffProp7}
\EF{ \exp\brak{-\tg f(X_T)} }{t} \\
=  \frac{1}{2\pi}\int_{\R} \exp\left[\phi\big(\tau,(0,-\beta+is)\big) \right.
\left. + \psi\big(\tau, (0,-\beta + is)\big) \cdot Y_t\right]\,\hat h_{\tg}(s)\,\,ds \,,%\tag{A.11}
\end{multline} 
where we need the dependence %of $\beta=\beta(\tg)$ and 
of $\hat h(s)=\hat h_{\tg}(s)$ on $\tg$.
Combining \eqref{AffProp6} and \eqref{AffProp7} yields
\begin{multline*}
\EF{ f(X_T) \exp\brak{-\tg f(X_T)} }{t} =   - \frac{\partial}{\partial \zeta} 
\left( \frac{1}{2\pi}\int_{\R} \exp\Big[\phi\big(\tau,(0,-\beta+is)\big) \right.\\
\left.+\,\,\psi\big(\tau, (0,-\beta+is)\big) \cdot Y_t\Big]\,\hat h_{\zeta}(s)\,\,ds \Bigg) 
\right|_{\zeta=\tg}\,.\tag*{\ensuremath{\square}}
\end{multline*}

\noindent {\bf Proof of Corollary \ref{AffPropCor}}\\
\noindent Expression \eqref{AffPropCorEQ} is an immediate consequence of \eqref{AffPropEQ} in Theorem \ref{MAINTHM}  
with $f(x)=x$,
and the fact that there is no
need of Fourier methods to compute the denominator $H(\tg)$ in the analogue to \eqref{AffProp2}
\begin{equation}\label{AffPropCor2}
S^1_t = \frac{\EF{X_T \exp(-\tg^1 X_T)}{t}}{\EF{\exp(-\tg^1 X_T)}{t}} \,,%\tag{A.12}
\end{equation}
since the affine transformation formula directly applies to the denominator in \eqref{AffPropCor2}.
We recall that
$\brak{T,(0,-\tg^1)} \in \mE$.
Now we only need to compute $\frac{\partial}{\partial \zeta}\EF{ \ef{-\zeta X_T} }{t}$, the actual derivative in formula \eqref{AffPropEQ}.
However, from \eqref{affinedef2} it follows that
\begin{equation*}
-\frac{\partial}{\partial \zeta}\EF{ \exp\brak{-\zeta X_T} }{t} 
=   \exp\edg{\phi\big(\tau,u\big) + \psi\big(\tau,u\big)\cdot Y_t} 
 \left.\big[ \partial_{u_x} \phi(\tau,u) + 
\partial_{u_x} \psi(\tau,u)\cdot Y_t \big] \right|_{u=(0,-\zeta)}\,.
\end{equation*}
Combining the above with \eqref{AffPropCor2} yields \begin{equation*} S^1_t = \left. \big[ \partial_{u_x} \phi(\tau,u) + 
\partial_{u_x} \psi(\tau,u)\cdot Y_t \big] \right|_{u=(0,-\tg^1)}\,.\end{equation*}%}_{=: G(\tau, Y_t)}.
As to the remaining securities $S^2,\dots,S^K$, their price processes given in \eqref{AffPropCorEQa} 
directly follow from formula \eqref{mainTHMform} in Theorem \ref{MAINTHM} and
the discussion above. \hfill\ensuremath{\square}\\

\noindent {\bf Proof of Theorem \ref{ThmEx1}}\\
\noindent An application of Theorem \ref{MAINTHM} with $\alpha^k = 0$, for all $k=1,\dots,N$, 
in addition to the observation that the Fourier transforms are all integrable functions yields the desired result.~As to the second
claim of integrability, straightforward calculations show that there exist constants $\hat M, \hat z > 0 $, just depending on the 
model parameters, which give
\begin{equation*}
\underset{f \in \{\hat g, \hat h, (\hat g^k)_{k=1}^N\}}{\max}\int_\R |f(s)| ds < \hat M \int_\R \frac{1}{s^2 + \hat z} ds < 
\infty\,. \tag*{\ensuremath{\square}}
\end{equation*}

\input{appendix02}

\noindent {\bf Proof of Theorem \ref{thm2101}}\\
\noindent By assumption, the conditions of Theorem \ref{density} are satisfied.%
~Recall that the equilibrium price is obtained by the change of measure from $P$ to $Q$, that is:
\begin{equation*}
S^k_t = \EFQ{S^k_T}{t} =\, \EFQ{f^k(X_1,\dots,X_N)}{t} 
=\, \EF{\frac{dQ}{dP}f^k(X_1,\dots,X_N)}{t} \EF{\frac{dQ}{dP}}{t}^{-1}\,.
\end{equation*}
By \eqref{density2}, we know that $\frac{dQ}{dP}$ is a function of $S_T$ and hence of $X_1,\dots,X_N$, which is given in \eqref{eq2105}.
Then we compute the regular conditional distribution of $\,(X_1,\dots,X_N)\,$ given $(\xi^1_t,\dots,\xi^N_t)$.
Using the independence of the market factors, the Markov property of $\xi$, the Bayes formula, and observing 
that, given $(X_1,\dots,X_N)=(x_1,\dots,x_N)$, $\xi^i_t$ is Gaussian with mean $\sigma_ix_i t$ 
and variance $\frac{tT}{T-t}$, yields \eqref{eq2106}.% and we are done. 
$\text{}$\hfill\ensuremath{\square}\\

\noindent {\bf Proof of Proposition \ref{prop3201}}\\
\noindent 
The integrability assumptions on $X$ together with \citep[Theorem 7.17]{LiptShir} yield that the innovation Brownian motion
$W_t$ in \eqref{eq2108} is well-defined for $t<T$.
By the Fujisaki-Kallianpur-Kunita Theorem, see \citep[Proposition 2.31]{BainCrisan},
both expressions appearing
in \eqref{eq2104a} allow for a representation with respect to $W$.
Furthermore, we even know the structure of the integrands.
Specifically, for every function $\varphi:\R\rightarrow\R$ such that $\varphi(X) \in L^2(P)$ and for $t<T$, we
obtain that
\begin{equation}\label{eq2109} 
\EF{\varphi(X)}{t} =  \E{\varphi(X)} + \int_0^t\frac{\sigma T}{T-u}V^\varphi_udW_u\,,%\tag{A.15}
\end{equation}
where $V^\varphi_t$, the conditional covariance of the market factor with the function $\varphi$, is given by 
\begin{equation}\label{covariance} 
V^\varphi_t = \EF{\varphi(X)X}{t} - \EF{\varphi(X)}{t}\EF{X}{t}\,,%\tag{A.16}
\end{equation}
as shown in \citep[Section V]{andrea04}. 
The dynamics \eqref{eq2111} then follow by \eqref{eq2109} in combination with \eqref{covariance} 
and an application of the It\^o product rule to \eqref{eq2104a}.\hfill\ensuremath{\square}\\

\noindent {\bf Proof of Corollary \ref{cor2101}}\\
\noindent The relation $\tg > \kappa - 1$ ensures that the assumptions of Theorem \ref{density} are met.%
~It remains to apply Theorem \ref{thm2101} and explicitly work out the integrals in 
\begin{equation*}
\frac{\int_0^\infty x \,(1/\kappa)\exp\brak{-x/\kappa} \exp\brak{-\tg x}\exp\edg{\frac{T}{T-t}\brak{\sigma x \xi_t -
\frac{1}{2}(\sigma x)^2 t}}dx}{\int_0^\infty (1/\kappa)\exp\brak{-x/\kappa} \exp\brak{-\tg x}\exp\edg{\frac{T}{T-t}\brak{\sigma x \xi_t -
\frac{1}{2}(\sigma x)^2 t}}dx}\,,
\end{equation*}
which is done by combining \citep[Section VII]{andrea04} and \eqref{eq2116},
resulting in formulae \eqref{eq2114} and \eqref{eq2115}.\hfill\ensuremath{\square}
\text{}\\

\section{Addendum to Section \ref{sec022}: Regular Affine Processes}\label{appendix011}
This proposition concerning the characterization of a regular affine process by its admissible parameters is stated without proof and we 
refer to \citep[Theorem 2.7]{duffieAffine} or  
\citep[Theorem 2.6 and Equations (2.2a),(2.2b)]{MKRphd} for two different approaches to prove it. 
\begin{prop}
Let $Y$ be a regular affine process with state space $D$.~Let $F$ and $R$ be as in Definition \ref{defaffreg}.~Then there exists a set of 
admissible parameters $(A,A^i,b,b^i,c,c^i,m,\mu^i)_{i=1,\dots,d}$ such that $F$ and $R$ are of the L\'evy-Khintchine form.
\begin{align}
F(u)& = \frac{1}{2} \langle u,Au \rangle + \langle b, u\rangle - c + \int_{\R^d\backslash\{0\}}\left(\ef{\langle\xi,u \rangle} - 1 - 
\langle h(\xi),u \rangle \right) m(d\xi)& \label{levykin1} \\
R_i(u)& = \frac{1}{2} \langle u,A^iu \rangle + \langle b^i, u\rangle - c^i + \int_{\R^d\backslash\{0\}}\left( \ef{\langle\xi,u \rangle} - 1 - 
\langle \chi^i(\xi),u \rangle \right) \mu^i(d\xi)\,,&\label{levykin2}
\end{align}
where $A,A^1,\dots,A^d$ are positive semi-definite real $d\times d$-matrices; $b,b^1,\dots,b^d$ are $\R^d$-valued vectors;
$c,c^1,\dots,c^d$ are positive non-negative numbers; $m$ and $\mu^1,\dots,\mu^d$ are L\'evy measures on $\R^d$, and finally $h$ and 
$\chi^1,\dots,\chi^d$
are suitably chosen truncation functions for the respective L\'evy measures.%
~Furthermore, the generator $\mathcal A$ of \,$Y$ is given by
\begin{alignat}{2}
\mathcal A \varphi(x) = \e\frac{1}{2}& \sum_{k,l=1}^d\brak{ A_{kl} + \sum_{i\in I} A^i_{kl} x_i }  
\frac{\partial^2 \varphi(x)}{\partial x_k\partial x_l} 
+ \langle b + \sum_{i=1}^d b^i x_i, \nabla\varphi(x)  \rangle - \brak{ c + \sum_{i\in I} c^ix_i } \varphi(x) \nonumber\\
+& \,\int_{D \backslash \{0\}} ( \varphi(c+\xi) -\varphi(x) - \langle h(\xi),  \nabla\varphi(x) \rangle ) m(d\xi) \nonumber\\
+&\,\sum_{i \in I} \int_{D \backslash \{0\}} 
( \varphi(c+\xi) -\varphi(x) - \langle \chi^i(\xi),  \nabla\varphi(x) \rangle ) x_i\mu^i(d\xi)\,,\label{generator}
\end{alignat}
and $\phi$, $\psi$ satisfy the following system of ODEs
\begin{align}
\partial_t \phi(t,u) &= F(\psi(t,u))\,,\q \phi(0,u)=0 \label{eqapp11}\\
\partial_t \psi(t,u) &= R(\psi(t,u))\,,\q \psi(0,u)=u \label{eqapp12}\,.
\end{align}
\end{prop}

%% file: appendix02.tex
\noindent {\bf Proof of Theorem \ref{maintheo}}\\
\noindent The process $Y = (V,X)$ belongs to a subclass of affine processes,
namely to the $\R^2$-valued affine diffusions.\footnote{We emphasize 
that we would not have needed the complete theory on general affine processes including various possible behavior of jumps, 
had we only considered pure diffusion processes, since it was shown in~\citep[Theorem 10.1]{Fili} 
that every diffusion Markov process with continuous diffusion matrix is affine, if and only if the functions $b$ and $\rho\rho^T$  
are affine in the state variable and the solutions $\phi$ and $\psi$ of the Riccati equations 
satisfy ${\rm Re}(\phi(t,u) + \psi(t,u)\cdot y)\le 0$, for all $y \in D$ and $(t,u) \in \R_{+} \times i \R^d$.~Our 
equilibrium approach can cover more sophisticated models than pure diffusions though.}
That is, $Y$ is a solution to the stochastic differential equation $dY_t = \mu(Y_t)dt + \rho(Y_t)dW_t$, with $Y_0 = y_0$,
for a continuous function $b:D \to \R^2$ and a measurable function $\rho:D \to \R^{2\times 2}$ such that $y \mapsto\rho(y)\rho(y)^T$ 
is continuous.
In particular, the set ${\rm int}\,\D_{0+}$ from Section \ref{ap} %\eqref{defanalytic1} 
is non-empty and thus the affine transorm formula can be extended.~See for instance 
the discussion on explosion times of the Heston model
in \cite{MomExplo}.~Furthermore, the process $Y$ is conservative and, hence, so is the stopped process $Y^T$.
Combining \eqref{generator} with the fact that the generator of $(V,X)$ is determined by its
diffusion matrix $\rho\rho^T$ and its drift vector $b$,
we identify the admissible parameters in \eqref{levykin1}, \eqref{levykin2} and \eqref{generator}, where the parts connected
with jumps do not play a role here.
Hence we conclude that the conditional characteristic function of $Y$ allows a representation as follows
\begin{equation}\label{affinetransHest}
\EF{\exp\brak{u\cdot Y_T}}{t} = \exp\edg{\phi(\tau,u) + \psi(\tau,u)\cdot Y_t}\,,%\tag{A.13}
\end{equation}%nd{align}
whenever $(T,u)=(T,(u_v,u_x)) \in \mE_{\C}$, so in particular for $(T,(u_v,u_x)) \in \mE$.~The functions $\phi$ and $\psi$ 
satisfy the following system of Riccati equations
\begin{align*}
\partial_t \phi(t,u)& = \kappa \,\psi_1 (t,u) + \mu\, \psi_2 (t,u)\,,& \phi (0,u)& =  0\\
\partial_t \psi_1 (t,u)& = \frac{1}{2} \sigma^2 {\psi_1 (t,u)}^2 - \lambda \psi_1 (t,u) + \frac{1}{2} 
{\psi_2 (t,u)}^2\,,&  \psi_1 (0,u)& = u_v\\
\partial_t \psi_2 (t,u)& = 0\,,&  \psi_2 (0,u)& = u_x\,.\tag{R}
\end{align*}
A solution to the above system (R), evaluated at the vector $u = (0,u_x)$, is 
given by\footnote{Compare \citep[Lemma 10.12]{Fili}.
For $u_x = \lambda/\sigma$ we 
set $\psi_1(t,(0,\frac{\lambda}{\sigma})) = t/(2+\lambda t)$, resembling the limit and still 
satisfying $\psi_1(0,(0,\lambda/\sigma))=0$.}
\begin{align*}
\phi \big(t,(0,u_x)\big)& = \frac{2\kappa}{\sigma^2}\log\left( \frac{2\theta(u_x)
\exp\brak{\frac{\theta(u_x)+\lambda}{2}t}}{\theta(u_x)(\ef{\theta(u_x)t}+1)+ 
\lambda(\ef{\theta(u_x)t}-1)}\right) + \mu u_xt\,,\\
\psi_1\big(t,(0,u_x)\big)& = \frac{u_x^2(\ef{\theta(u_x)t}-1)}{\theta(u_x)(\ef{\theta(u_x)t}+1) + \lambda(\ef{\theta(u_x)t}-1)}\,,\nonumber\\
\psi_2\big(t,(0,u_x)\big)& = u_x\,.
\end{align*}
where
\begin{equation*}
\theta(u_x) =  \left\{\begin{array}{ll}\sqrt{\lambda^2 - \sigma^2\,u_x^2} \q&\mbox{if}\q \abs{u_x} <
\frac{\lambda}{\sigma} \\i \sqrt{ \sigma^2\,u_x^2 - \lambda^2 } \q&\mbox{if}\q \abs{u_x} >
\frac{\lambda}{\sigma}\end{array}\right. \,.\label{sqrt}
\end{equation*}
Following \cite{MomExplo} and recalling that $\lambda>0$, we distinguish two different cases
\begin{align*}
t^+(u_x) = \left\{\begin{array}{lc}+\infty\q & \abs{u_x} <\frac{\lambda}{\sigma} \\
\frac{2}{|\theta(u_x)|}\left( \arctan\frac{|\theta(u_x)|}{-\lambda}  + \pi  \right) & \abs{u_x} >
\frac{\lambda}{\sigma}\end{array}\right.
\end{align*}
such that $(T,(0,u_x)) \in \mE \subseteq\mE_{\C}$, 
for all $T \le t^+(u_x)$.\footnote{Basically, this is exactly the time interval on which the solutions of the Riccati
equations do not explode.}~Hence, as long as $T < t^+(u_x)$, formula \eqref{affinetransHest} holds for all $u=(0,u_x)$, where $u_x \in \R$.
It now follows from \eqref{AffPropCorEQ} in Corollary \ref{AffPropCor} that, for all $t\in[0,T]$,
\begin{equation}\label{app02eq01}
S_t =
\left.\big[ \partial_{u_x} \phi(\tau,u) + \partial_{u_x} \psi_1(\tau,u) V_t  + 
\partial_{u_x} \psi_2(\tau,u)X_t \big]\right|_{u=(0,-\gamma)} \,,%\tag{A.14}
\end{equation}
Next we need to compute the derivatives of $\phi(t,u)$ and $\psi(t,u)$ with respect to $u_x$.%
~Of course we have $\partial_{u_x} \psi_2(\tau,u)\equiv 1$
and a straightforward calculation yields, 
with $\theta:= \theta(-\gamma)$ and $\theta':= [\partial_{u_x}\theta](-\gamma)$,
\begin{equation*}
\partial_{u_x} \phi(\tau,(0,-\gamma)) = T(\tau,\gamma)\q\mbox{and }\q \partial_{u_x} \psi_1(\tau,(0-\gamma)) 
= - \gamma \varGamma(\tau,\gamma)\,.
\end{equation*}
This, together with \eqref{app02eq01}, is \eqref{maintheo1}, the proof is complete.\hfill\ensuremath{\square}\\